%% file: ISIT2020_v2.tex
\documentclass[conference,a4paper]{IEEEtran}
\usepackage{cite}
\usepackage{graphicx,color,epsfig,rotating}
\usepackage{amsfonts,amsmath,amssymb,bbm}
\usepackage{algorithm}
\usepackage{algpseudocode}
\usepackage{subfigure}
\usepackage{amsmath}
\usepackage{cite}
\usepackage{mdwtab}
\usepackage{placeins}
\usepackage{psfrag, graphicx}
\usepackage[latin1]{inputenc}
\usepackage{amssymb}
\usepackage{multirow}
\usepackage{stfloats}
\usepackage{tabularx}
\usepackage{booktabs}
\usepackage{url}
\usepackage{bm}

\input{macros}

\newtheorem{defn}{Definition}
\newtheorem{example}{Example}
\newtheorem{theorem}{Theorem}
\newtheorem{lemma}{Lemma}
\newtheorem{remark}{Remark}
\makeatletter
\newcommand{\subalign}[1]{%
  \vcenter{%
    \Let@ \restore@math@cr \default@tag
    \baselineskip\fontdimen10 \scriptfont\tw@
    \advance\baselineskip\fontdimen12 \scriptfont\tw@
    \lineskip\thr@@\fontdimen8 \scriptfont\thr@@
    \lineskiplimit\lineskip
    \ialign{\hfil$\m@th\scriptstyle##$&$\m@th\scriptstyle{}##$\crcr
      #1\crcr
    }%
  }
}
\makeatother

\ifodd 1
\newcommand{\mj}[1]{{\color{red} #1}}
\else
\newcommand{\mj}[1]{#1}
\fi

\ifodd 1
\newcommand{\xz}[1]{{\color{blue} #1}}
\else
\newcommand{\xz}[1]{#1}
\fi

\begin{document}
\sloppy
\title{A New Design Framework on D2D Coded Caching with Optimal Rate and Less Subpacketizations}
\author{
\IEEEauthorblockN{Xiang Zhang, Mingyue Ji }
	\IEEEauthorblockA{Department of Electrical and Computer Engineering, University of Utah\\
		Salt Lake City, UT, USA\\
		Email: \{xiang.zhang, mingyue.ji\}@utah.edu}}
\maketitle

\begin{abstract}
In this paper, we propose a new design framework on Device-to-Device (D2D) coded caching networks with optimal communication load (rate) but significantly less file subpacketizations compared to that of the well-known D2D coded caching scheme proposed by Ji, Caire and Molisch (JCM). The proposed design framework is referred to as the {\em Packet Type-based (PTB) design}, where each file is partitioned into packets according to their  pre-defined types while the cache placement and user multicast grouping are based on the packet types. This leads to the so-called {\em raw packet saving gain} for the subpacketization levels. By a careful selection of transmitters within each multicasting group, a so-called {\em further splitting ratio gain} of the subpacketizatios can also be achieved. By the joint effect of the {\em raw packet saving gain} and the {\em further splitting ratio gain}, an order-wise subpacketization reduction can be achieved compared to the JCM scheme while preserving the optimal rate. 
In addition, as the first time presented in the literature according to our knowledge, we find that unequal subpacketizaton is a key to achieve subpacketization reductions when the number of users is odd. As a by-product, instead of directly translating shared link caching schemes to D2D caching schemes, at least for the sake of subpackeitzation, a new design framework is indeed needed.

\end{abstract}
\section{Introduction}
\label{section: intro}
Coded caching has been shown to be an efficient approach to handle dramatically increased traffic in the current Internet. In \cite{maddah2014fundamental}, Maddah-Ali and Niesen (MAN) introduced a centralized shared-link caching network model, where a central controller serves $K$ users, each of which is equipped with a cache of size $M$ files from a library of $N$ files, via an errorless broadcast link (shared link). In order to achieve the optimal worst-case rate under uncoded cache placement,\footnote{Rate is defined as the total number of file transmissions in the network.} a cache placement and a coded delivery scheme were proposed \cite{maddah2014fundamental} and required to partition each file into ${K \choose t}$ packets where $t = \frac{KM}{N}\in\mathbb{Z}^+$. Later, \cite{yan2017placement} shows that this file subpacketization level is necessary to achieve the optimal rate under a so-called Placement Delivery Arrary (PDA) design based on uncoded cache placement. In order to reduce the subpacketization level, the authors in \cite{Shanmugam2016,shanmugam2017isit,shangguan2018centralized,tang2018coded,jin2019caching} proposed schemes based on various combinatorial designs and showed that the subpacketization can be reduced at a cost of a higher transmission rate (or higher traffic load). Ji, Molisch and Caire (JCM) extended the shared-link caching model to Device-to-Device (D2D) coded caching networks, where no central controller is present and all users serve each other via individual shared links \cite{ji2016fundamental}. This network model has been studied extensively in the literature and a few examples of the information-theoretic study are given in \cite{Sengupta15isit, 8491204,8830435,wan2019device}. Under uncoded cache placement, \cite{ji2016fundamental} proposed a caching scheme referred to as the {\em JCM scheme} that achieves the optimal worst-case rate of $R(M)=\frac{N}{M}\left(1-\frac{M}{N}\right)$ when $N\geq K$. In this case, $R(M)$ is surprisingly not a function of $K$ and hence it is scalable. In order to achieve this rate, the required number of packets (\emph{subpacketization level}) is
$F^{\rm JCM}=t\binom{K}{t}$, which can be impractical for large $K$. Efforts have been made in reducing the subpacketization levels for the D2D coded caching problem \cite{woolsey2017caching, wang2017placement, woolsey2017towards, woolsey2018caching}.   
For example, a design approach named {\em D2D placement delivery array} (DPDA) was introduced in \cite{wang2017placement}, which designed new DPDA schemes when $t=2, t=K-2$, for which the JCM scheme is actually not optimal in terms of subpacketization although it achieves the optimal rate. 

In this paper, we propose a new design framework called {\em Packet Type-based (PTB) design} tailored for subpacketization reduction in D2D coded caching while preserving the optimal rate. Specifically, in the PTB design, D2D users (or nodes) are first partitioned into multiple groups. Then packet types are designed based on the user grouping and different types of multicasting groups are designed based on the packet types. {Note that the JCM scheme contains all the packet types and all multicasting group types; any $t+1$ users form a multicasting group and every user in each multicasting group transmit ``symmetric" coded multicast message that is useful to all the group members.} In contrast, the proposed PTB scheme excludes certain packet types which leads to a reduction of the subpacketization level. This is referred to as the {\em raw packet saving gain}. In addition, in each multicasting group, it is possible that not all nodes will perform as transmitters. Hence, based on a careful selection of the transmitters within each type of multicasting group, a so-called {\em further splitting ratio gain} can also be obtained. While preserving the optimal rate, the joint effect of the {\em raw packet saving gain} and the {\em further splitting ratio gain} can lead to an order-wise subpacketization reduction compared to the JCM scheme, where none of these gains is available. In fact, the PTB design problem can be cast into an integer optimization problem subject to node cache memory constraints and the design variables are the choices of possible transmitters within each multicasting group type. {Moreover, according to our knowledge, it is the first time in the literature showing that unequal subpacketizaton is one of the keys to achieve a subpacketization gain when $K$ is odd.} 

In \cite{ji2016fundamental}, in order to achieve the optimal rate, the JCM scheme proposed a direct translation from MAN scheme \cite{maddah2014fundamental} by splitting each packet further into $t$ packets. It turns out that when the cache placement is uncoded and the delivery scheme is one-shot,\footnote{{ One-shot delivery scheme means that for each coded transmission, every user can successfully decode one requested packet.}} this translation holds in general and it seems that the design procedure for the D2D coded caching scheme should be that 1) designing a shared-link coded caching scheme; 2) translating it into D2D coded caching scheme.
As a by-product of the PTB design, we show that the above design methodology is not optimal in terms of subpacketization in general. Hence, in order to achieve good subpackeitzations in D2D coded caching networks, a new design framework is indeed needed. Due to the limitation of space, all the proofs and detailed descriptions 
can be found in \cite{zhang2018D2Dlong}.
\paragraph*{Notation Convention}
$|\cdot|$ represents the cardinality of a set. 
$\mathbb{Z}^+$ denotes the non-negative integer set and $\mathbb{Q}^+$ denotes the set of positive rational numbers. $[n]:=\{1,\cdots,n\}$, 
$[m:n]:=\{m,m+1,\cdots,n\}$ for some $m<n$ and ${\bf a}^n = (\underbrace{a, a, a, \cdots a}_{n\; {\rm terms}})$.

\section{Problem Formulation and Illustrations}
\label{sec: problem formulation}
\subsection{General Problem Description}
Consider a D2D caching network with a user set $\Uc$ where $|\Uc| = K$. Each file from a library $\{W_n: n \in [N]\}$ of $N$ files consists of $F$ packets with equal length.\footnote{{ The equal length assumption is for the ease of presentation. In practice, the length of each packet may be a design parameter since only the file length should be fixed. In the main results, we will see that if this assumption is relaxed, it is possible to achieve additional gain in terms of the subpacketizations.}} The system operates in two separate phases, {\em i.e.}, the \emph{cache placement phase} and \emph{delivery phase} as described in \cite{ji2016fundamental}. In the cache placement phase, each user $k$ stores up to $MF$ packets from the file library. This phase is done without the knowledge of the users' requests. In the delivery phase, each user $k$ reveals its request for a specific file $W_{d_k},d_k\in[N]$ to other users. Let $\mathbf{d}:=(d_1,d_2,\cdots,d_K)$ denote the user demand vector. Since users have already cached some of the files, the task in the delivery phase is to design a corresponding transmission scheme for each user based on the cache placement and the user demand vector so that the users' demands can be satisfied with vanishing error probability. The objective is to minimize the transmission rate (communication load) defined as the total number of transmitted bits normalized by the file size. 
In this paper, our goal is to propose a new design framework based on combinatorial optimizations such that the subpacketization level of each file is significantly reduced while preserving the optimal rate. In the rest of this paper, we use $F^{\rm JCM}$ and $F^{\rm PTB}$ to represent the subpacketization level of the JCM scheme and the proposed PTB scheme.

\subsection{JCM D2D Coded Caching Scheme}
The cache placement in JCM scheme is the same as MAN cache placement scheme introduced in \cite{maddah2014fundamental}. Let $t:=\frac{MK}{N}$, each file $W_n$ is divided into ${K \choose t}$ disjoint sub-files denoted by $W_{n, \Tc}$, where $\Tc \subset [K]$ and $|\Tc| = t$. The size of each sub-file is $F/{K \choose t}$ packets. Each user $u$ caches all the packets in sub-files $W_{n, \Tc}$, for all $n \in [N]$ and $u \in \Tc$.

In the delivery phase, for each multicasting group with $t+1$ users, it can be seen that there exists a unique sub-file that is available at $t$ users and is requested by the  remaining  user. In order to symmetrize all transmissions (minimize the communication load), each sub-file needs be divided into $t$ packets, where $W_{n, \Tc} = \{W_{n, \Tc}^{(j)}\}$, where $j \in [t]$. It can be seen that each file needs to be partitioned into $t{K \choose t}$ packets. Hence, the transmitted coded multicast message for user $u$ is given by
\be
Y_{\Ac^u}^u = \bigoplus_{k \in \Ac^u} W_{d_k, \{\Ac^u \cup \{u\} \setminus \{k\}\}}^{(j)},
\ee
for all the user groups $\Ac^u \subset [K] \setminus \{u\}$ with $|\Ac^u| = t$ and for $j \in [t]$ that is not chosen yet. It can be seen that all users can decode all the desired sub-files and the transmission rate is given by $R = \frac{N}{M}(1-\frac{M}{N})$, which is optimal if $N > K$.

\subsection{An Example to Illustrate PTB schemes}
\label{sec: example} 
\begin{figure}
\begin{center}
\includegraphics[width=0.45\textwidth]{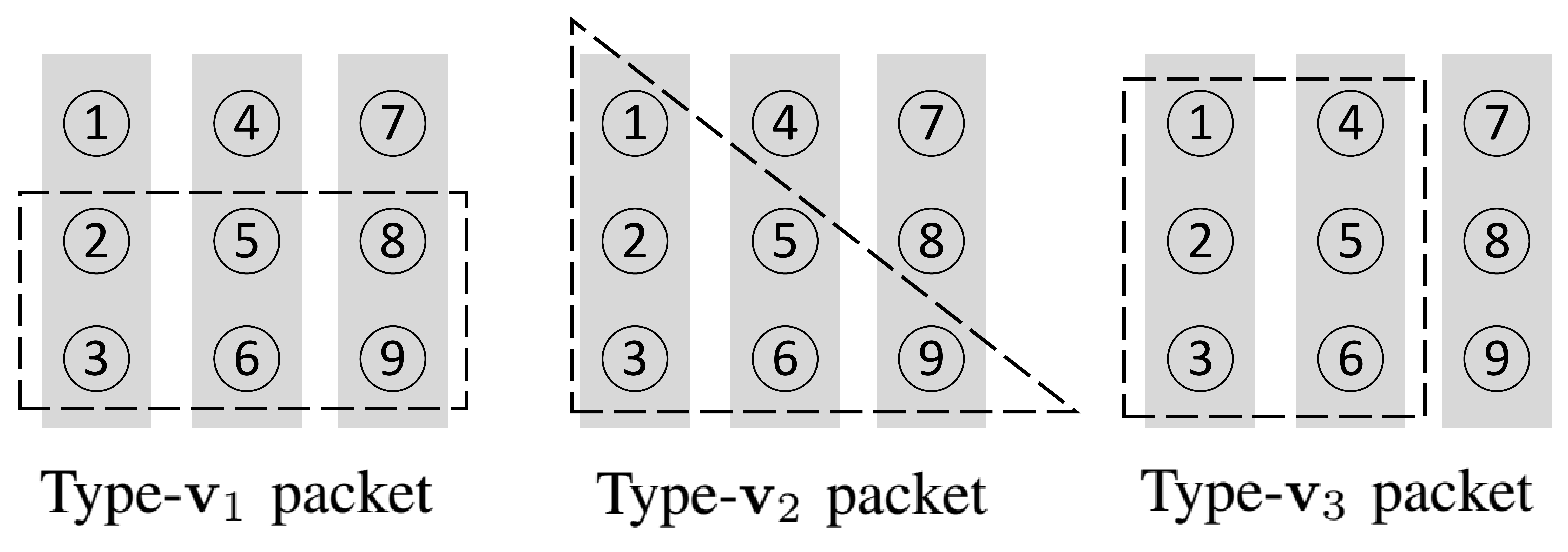}
\caption{An illustration of packet types under node grouping $\mathbf{q}=(3,3,3)$ with three specific groups $\Qc_1=\{1,2,3\},\Qc_2=\{4,5,6\}$ and $\Qc_3=\{7,8,9\}$.}
\label{fig: packet type}
\end{center}
\vspace{-0.5cm}
\end{figure}

In this section, we present an example to show the key ideas of the PTB design approach and its difference from the JCM scheme. We consider a D2D coded caching network with parameters $K=9,N=3,M=2$ and $t=KM/N=6$. The user set is $\Uc = [9]$. We evenly partition $\Uc$ to $3$ groups, denoted as $\Qc_1, \Qc_2$ and $\Qc_3$, where $\Qc_1 = \{1,2,3\}$, $\Qc_2 = \{4,5,6\}$ and $\Qc_3 = \{7,8,9\}$ (see Fig.~\ref{fig: packet type}). We use a {partition vector} $\qv = (3,3,3)$ to indicate such a node grouping, where each element in $\qv$ denotes the number of nodes in each group. One of the key ideas of the proposed approach is to introduce the {\em packet type}, which is a partition of $t=6$ nodes from all user groups. In this example, there are three partition vectors (i.e, three packet types) $\vv_1 = (2,2,2)$, $\vv_2=(3,2,1)$ and $\vv_3 = (3,3,0)$, where the sum of all the elements in each partition vector is 6. The packets of type $\vv_1$ are cached by $2$ users in $\Qc_1$, $2$ users in $\Qc_2$ and $2$ users in $\Qc_3$ respectively. The meaning of packet types $\vv_2$ and $\vv_3$ follows similarly. Fig.~\ref{fig: packet type} shows an example for all three packet types. For instance, for packet type $\vv_1$, Fig.~\ref{fig: packet type} illustrates the case where the packets are cached by users $\Tc = \{2,3,5,6,8,9\}$. When all packets have the same size, in order to design the cache placement, we need to decide the number of packets for each type. It can be shown that packet type $\vv_3$ can be \emph{excluded}. This means that only the packet types $\vv_1$ and $\vv_2$ are needed for designing the D2D coded caching scheme. This results in ${3 \choose 2}^3 + {3 \choose 1} {3 \choose 2} {2 \choose 1} {3 \choose 1} = 81$ sub-files while the JCM scheme requires ${9 \choose 6} = 84$. At this stage, the saving of the PTB design in terms of sub-files is $84-81=3$, which is referred to as the {\em  raw packet saving gain}. It can also be seen that the JCM scheme includes all possible packet types $\vv_1, \vv_2$, and $\vv_3$. \footnote{{ Note that the subpacketization is not final yet, we call the unit of the file partition as ``sub-files", not packet, which is the smallest unit in a file.}}
Moreover, each sub-file of type $\vv_1$ needs to be partitioned into $4$ packets and each sub-file of type $\vv_2$ needs to be partitioned into $3$ packets according to the PTB scheme. Note that each sub-file in the JCM scheme needs to be partitioned into $t=6$ packets. This further reduction of subpacketization of PTB scheme is referred to as the {\em further splitting ratio gain}. 
Hence, using the PTB design, we can compute the number of total packets needed for each file as ${3 \choose 2}^3 \cdot 4 + {3 \choose 1} {3 \choose 2} {2 \choose 1} {3 \choose 1} \cdot 3= 270$ while the number of packet per file required by JCM is $t{K \choose t} = 6{9 \choose 6} = 504$. Clearly, the cache placement, determined by the packet type, is that node $k$ stores any packet $W_{n,\mathcal{T}}^{(j)}$  for $k\in\mathcal{T}$, $j \in [4]$ $n \in [N]$. For example, Fig.\ref{fig: packet type} shows the four packets $\{W_{n,\mathcal{T}}^{[j]}:j\in[4]\}$ derived from the type $\vv_1$ sub-file $W_{n,\Tc}$ where $\Tc = \{2,3,5,6,8,9\}$.

In the delivery phase, the JCM scheme exploits the ``symmetric" multicasting group structure , where any $t+1$ out of $K$ users can form a multicasting group and any node in each multicasting group is a transmitter. For the proposed PTB scheme, this ``symmetry" breaks, which means that the multicasting groups have to be designed according to the packet types and the transmitter in each multicasting group has to be designated specifically. This means that we can have different ``types" of multicasting group. In this example, we have two types of multicasting groups denoted as $\sv_1 = (3,3,1^*)$ and $\sv_2=(3,2^*,2^*)$, where each element in $\sv_1$ and $\sv_2$ means the number of users from the corresponding user group and the symbol ``$*$" means that all nodes in the corresponding user group are selected as transmitters. For example, for a multicasting group $\Sc=[7]$ of type $\sv_1$, all users in $\Qc_1$ and $\Qc_2$ are included and one node $7$ from $\Qc_3$ is included. The only transmitter in this multicasting group is node 7 from $\Qc_3$. Note that the JCM scheme includes all types of multicasting groups in general and every node in each multicasting group is a transmitter. 

Next, we will illustrate the design of the coded multicast message. For a type-$\mathbf{s}_1$ multicasting group $\mathcal{S}_1=[7]$, node 7 is the only transmitter and it transmits three coded multicast messages $\bigoplus_{k\in[6]}W_{d_k,\mathcal{S}_1\backslash \{k\}}^{(j)}$, $j=1,2,3$ to other nodes in $\mathcal{S}_1$. Each node $k$ recovers its desired packets $\{W_{d_k,\mathcal{S}_1\backslash \{k\}}^{(j)}:j=1,2,3\}$ with the help of the cached packets while node 7 itself only transmits but receives nothing. For a type-$\mathbf{s}_2$ multicasting group $\mathcal{S}_2=[9]\backslash\{6,9\}$, the set of type-$\vv_1$ and $\vv_2$ subpackets involved are $\{W_{d_k,\mathcal{S}_2\backslash\{k\}}^{(j)}:j\in[4],k\in[3]\}$ and $\{W_{d_k,\mathcal{S}_2\backslash\{k\}}^{(j)}:j\in[3],k\in \mathcal{S}_2\backslash\mathcal{Q}_1\}$ respectively. Denote $W^{(j)}:=\bigoplus_{k\in[3]}W_{d_k,\mathcal{S}_2\backslash\{k\}}^{(j)},j\in[4]$. Each node $k\in\{4,5,7,8\}$ sends a coded multicast message $Y_k$ as follows.
\begin{eqnarray}
Y_4&=&W^{(1)}\oplus W_{d_5,\mathcal{S}_2\backslash\{5\}}^{(1)}\oplus W_{d_7,\mathcal{S}_2\backslash\{7\}}^{(1)} \oplus W_{d_8,\mathcal{S}_2\backslash\{8\}}^{(1)}\nonumber\\
Y_5&=&W^{(2)}\oplus W_{d_4,\mathcal{S}_2\backslash\{4\}}^{(1)}\oplus W_{d_7,\mathcal{S}_2\backslash\{7\}}^{(2)} \oplus W_{d_8,\mathcal{S}_2\backslash\{8\}}^{(2)}\nonumber\\
Y_7&=&W^{(3)}\oplus W_{d_4,\mathcal{S}_2\backslash\{4\}}^{(2)} \oplus W_{d_5,\mathcal{S}_2\backslash\{5\}}^{(2)} \oplus W_{d_8,\mathcal{S}_2\backslash\{8\}}^{(3)}\nonumber\\
Y_8&=&W^{(4)}\oplus  W_{d_4,\mathcal{S}_2\backslash\{4\}}^{(3)}\oplus W_{d_5,\mathcal{S}_2\backslash\{5\}}^{(3)}\oplus W_{d_7,\mathcal{S}_2\backslash\{7\}}^{(3)} \nonumber
\end{eqnarray}
from which we can see that all nodes can recover their desired packets. Since each coded message is simultaneously useful for $t=6$ nodes, the transmission rate is $\frac{N}{M}-1$. The transmission procedure for other multicasting groups is similar.

In the next section, we will generalize the proposed scheme in this example and present the main results of this paper. 

\subsection{General Packet Type Based (PTB) Design Framework}
The specific design proposed in the previous example is not unique and can be generalized by solving the following optimization problem, where the solutions are called PTB designs. 
\vspace{-0.1cm}
\begin{eqnarray}
&F^{\rm PTB}= \min_{\bm{\alpha}^{\rm LCM}, \Fm}& \bm{\alpha}^{\rm LCM}\mathbf{F}^{\rm T}\label{eq: optimization 1} \\
&{\rm s.t.}&\bm{\alpha}^{\rm LCM}\in \Phi, \label{eq: optimization 2} \\ 
&& \bm{\alpha}^{\rm LCM}\Delta\mathbf{F}_i^{\rm T}=0, i\in[N_{ d}-1]\label{eq: optimization 3}
\end{eqnarray}
where $\mathbf{F}^{\rm T}$ denotes the {\em raw packet number vector}; $\bm{\alpha}^{\rm LCM}$ denotes the {\em further splitting vector} and all the notations will be explained in detail in the Appendix. It can be seen that this is an integer optimization problem and cannot be solved analytically in general. In the following, we will provide some solutions in some specific parameter regimes. Note that the solutions of this combinatorial problem achieve the transmission rate $\frac{N}{M} (1-\frac{M}{N})$, which is optimal when $K>N$.

\section{Main Results}
\label{section: main result}

In this section, we present the main theorems. First, we consider the scenario when $K$ is even and $t$ is large. 

\begin{theorem}
\label{theorem_1_large_t}
For even $\bar{t}:=K-t$, where $K=2m$, using the PTB design framework, the rate of $\frac{N}{M}-1$ in D2D coded caching networks is achievable and 
\vspace{-0.1cm}
\begin{eqnarray}
\frac{F^{\rm PTB}}{F^{\rm JCM}}=\Theta\left(\frac{f(\bar{t})}{K-\bar{t}}\right),
\end{eqnarray}
where $f(\bar{t}):=\prod_{i=1}^{\frac{\bar{t}}{2}}(2i-1)$ is a function which depends only on $\bar{t}$. Moreover, $\forall K \geq 2\bar{t}$, and $\bar t = O(\log\log K)$, $\frac{F}{F^{\rm JCM}}$ vanishes as $K$ goes to infinity. \hfill $\square$
\end{theorem}

From Theorem \ref{theorem_1_large_t}, it can be seen that when $K \leq 2  t$ is even and $t$ is large enough ({\em i.e.}, $t = K - O(\log\log K)$), an order gain in terms of subpacketization can be obtained using the PTB design compared to the JCM scheme while preserving the optimal rate. However, it can be seen that for small $t$, the PTB design achieving Theorem \ref{theorem_1_large_t} may result in an even worse subpacketization compared to the JCM scheme. In the following theorem, we provide a general result for even $K$ and $t$ based on a specific PTB design demonstrating subpacketization gains compared to the JCM scheme when $t$ is small.

\begin{theorem}
\label{theorem_2_groups}
For $(K,t)=(2q,2r)$ with $q\geq t+1$ and $r\geq 1 (r\in \mathbbm{Z}^+)$, using the PTB design framework with the two-group equal grouping, {\em i.e.}, $\mathbf{q}=(\frac{K}{2},\frac{K}{2})$, the rate of $\frac{N}{M}-1$ of D2D coded caching networks is achievable by the further splitting ratio vector $\bm{\alpha}^{\rm LCM}=(0,1,2,\cdots,r)$. Further, when $r \geq 2$, we have $\frac{F^{\rm PTB}}{F^{\rm JCM }}< \frac{1}{2}\left(1-\frac{1}{2^{t-1}}\right) $.
\hfill $\square$
\end{theorem}
 
When $K$ is odd, it is surprisingly more difficult than the case of even $K$. In Section~\ref{sec: example}, we provided an example showing that when $K=3m$ and $t = K-3$, $m \in \mathbb{Z}^+$, it is possible to exploit an equal user grouping to achieve an order gain of subpacketization level compared to the JCM scheme. However, in general, we may need to use the general PTB design framework (see \cite{zhang2018D2Dlong}) that exploits the heterogeneous packet size.

\begin{theorem}
\label{theorem_heter_pkt_size}
For $(K,t)=(2q+1,2r)$ with $q\geq 2r+1,r\geq 1 (r\in\mathbbm{Z}^+)$, using the two-group unequal grouping $\mathbf{q}=(\frac{K+1}{2},\frac{K-1}{2})$, the rate of $\frac{N}{M}-1$ in D2D coded caching networks is achievable by the further splitting ratio vector $\bm{\alpha}^{\rm LCM}=(0,2,4,\cdots,t-2,t,t,\cdots, t)$ and we have
\vspace{-0.1cm} \be \frac{F^{\rm PTB}}{F^{\rm JCM}}< \frac{1}{t}\left(\frac{\binom{t}{r}}{2^t}-1\right)+1\ee \hfill $\square$\vspace{-0.1cm}
\end{theorem}

From Theorem \ref{theorem_heter_pkt_size}, it can be seen that by using the general PTB design framework with the consideration of heterogeneous subpacket size, when $K$ is odd, a constant gain in terms of subpacketization compared to the JCM scheme can be achieved while preserving the optimal rate when $t$ small.

\section{Conclusions}
This paper proposed a new design framework called PTB designs for D2D coded caching networks. The key ideas of this specific design are 1) classifying packets by the packet types;   2) grouping users into multicasting groups based on the cached packet types and 3) asymmetrically assigning transmitters in each multicasting group. This new design approach is completely different from the original coded D2D caching schemes and shows that the coded D2D caching problem may need a new design idea in contrast to firstly designing a centralized coded caching problem and then convert it to a D2D coded caching scheme.

\appendices

\section{General Packet Type-based (PTB) Design Framework}
We will present the general PTB design framework by decomposing it into the concepts including {\em Node Grouping}, {\em Packet Type}, {\em Multicasting Group Type}, {\em Further Splitting Ratio (FSR)}, {\em Further Splitting Ratio Table (FSRT)}, {\em Memory Constraint Table (MCT)} and {\em PTB Design as an Integer Optimization Problem}.
\subsubsection{Node Grouping} The user set $\Uc$ is partitioned into $m\in\mathbb{Z}^+$ non-empty groups denoted by $\Qc_1,\Qc_2,\cdots,\Qc_m$, where the $i$-th group contains $|\Qc_i| = q_i$ nodes. We use a {\em partition vector} $\mathbf{q}:=({q_1,q_2,\cdots,q_m,0,0,\cdots,0})(|\qv|=K)$ to represent such a node grouping, satisfying $\sum_{i=1}^mq_i=K$ and $q_1\geq q_2\geq \cdots\geq q_m>0$. For a specific partition $\qv$, there are actually multiple ways to assign the set of $K$ nodes, but they are all considered the same partition/grouping. The number of groups $m$ and the number of nodes contained in each group, \emph{i.e.}, $\{q_i\}_{i\in[m]}$ are parameters to be designed. Let $N_d$ denote the number of distinct elements/parts\footnote{{ The partition vector $\mathbf{q}$ is a representation of partition of the integer $K$ in number theory, where each $q_i$ is called a \emph{part}. `0' is not considered as a part.}} in $\mathbf{q}$. We define a \emph{unique group} as the union of the non-empty groups containing the same number of nodes. The $i$-th ($i\in[N_d]$) {\em unique group}, denoted by $\mathcal{U}_i$, contains $\psi_i$ groups and each of these groups contains $\beta_i$ nodes, {\em i.e.}, $|\Uc_i|=\psi_i\beta_i$. It is clear that $\sum_{i=1}^{N_d}\beta_i\psi_i=K $ and $\sum_{i=1}^{N_d}\psi_i=m$. For example, let $K=7$ and the user set be $\Uc = [K]$. $\mathbf{q}=(3,2,1,1,0,0)$ is a partition vector representing a partition of $\Uc$ into $m=4$ groups which are $\Qc_1=\{1,2,3\}$, $\Qc_2=\{4,5\}$, $\Qc_3=\{6\}$ and $\Qc_4=\{7\}$. In this case, there are $N_{d}=3$ unique groups, {\em i.e.}, $\Uc_1 = \Qc_1= \{1,2,3\}$, $\Uc_2 = \Qc_2 = \{4,5\}$, $\Uc_3 =\Qc_3\cup \Qc_4 =\{6,7\}$. According to the definitions, we  also have $(\beta_1,\psi_1)=(3,1),(\beta_2,\psi_2)=(2,1),(\beta_3,\psi_3)=(1,2)$. Using the definition of unique groups, we can represent a partition vector $\qv =(\underbrace{\beta_1,\cdots,\beta_1}_{ \psi_1 \; {\rm terms}},\underbrace{\beta_2,\cdots,\beta_2}_{ \psi_2 \; {\rm terms}},\cdots,\underbrace{\beta_m,\cdots,\beta_m}_{ \psi_m \; {\rm terms}},0,\cdots,0   ) $ by a more compact form $\qv =(\bm{\beta}_1^{(\psi_1)},\bm{\beta}_2^{(\psi_2)},\cdots,\bm{\beta}_m^{(\psi_m)},\mathbf{0})$. Moreover, we call a node grouping an \emph{equal grouping} if all the groups contain the same number of nodes, {\em i.e.}, $q_1=q_2=\cdots=q_m=\frac{K}{m}$. Otherwise, it is called an \emph{unequal grouping}. Clearly, the example presented in Section \ref{sec: example} uses an unequal grouping.
\subsubsection{Packet Type} A \emph{packet type} refers to a partition of $t:=\frac{KM}{N}\in\mathbb{Z}^+$ nodes and is represented by a partition vector $\mathbf{v}:=(v_1,v_2,\cdots,v_t)$ satisfying $\sum_{i=1}^{t}v_i=t$ and $v_1\geq v_2\geq \cdots\geq v_t\geq 0$. Different partitions of $t$ correspond to different packet types. A \emph{raw packet} $W_{n,\mathcal{T}}$, for some $\Tc \subset \Uc$, $|\Tc| = t$ refers to a sub-file that is cached exclusively by a set of nodes in $\mathcal{T}$. Each packet type may contain multiple raw packets. Since not all packets types can appear under a given node grouping, we can exclude some invalid packet types, meaning that these packet types will not be used in the PTB design. This is called {\em raw packet saving gain}. In the delivery phase, raw packets (i.e, sub-files) might be further split into multiple \emph{packets}, {\em i.e.}, $W_{n,\mathcal{T}}=\{W_{n,\mathcal{T}}^{(i)}\}_{i\in[\alpha(\vv)]}$ where $\vv$ is the packet type and $\alpha(\vv)$ is called \emph{further splitting ratio}. 
Raw packets of the same type must have the same further splitting ratio. Note that all raw packets have the same further splitting ratio $\alpha(\vv)=t$, for any packet type $\vv$ in the JCM scheme.
\if{0}
{
\begin{example}
\label{example_illu_pkt_type}\textbf{(Packet Type)}
For $(K,t)=(6,3)$ and $\Uc = [K]$, consider the node grouping $\mathbf{q}=(3,3)$\footnote{For simplicity, we ignored the zeros in the partition vector $\qv$. The complete form should be $\mathbf{q}=(3,3,0,0,0,0)$.} with a specific node assignment $\mathcal{Q}_1=\{1,2,3\},\mathcal{Q}_2=\{4,5,6\}$, which is shown in Fig. \ref{figure_illu_pkt_types}. There are two different types of packets, {\em i.e.}, $\mathbf{v}_1=(3,0)$, meaning picking three nodes from either one of the two groups, and $\mathbf{v}_2=(2,1)$, meaning picking two nodes from one group and one node from the other group. For example, the packet $W_{n,\{4,5,6\}}$ which is cached in 
nodes 4, 5, 6 is a type-$\mathbf{v}_1$ packet. The packet $W_{n,\{3,5,6\}}$ which is cached in 
nodes 3, 5, 6 is a type-$\mathbf{v}_2$ packet. It can be seen that 
there are in total $2\binom{3}{3}=2$ type-$\mathbf{v}_1$ packets and $2\binom{3}{2}\binom{3}{1}=18$ type-$\mathbf{v}_2$ packets. These two packets are called raw packets since they have not been further split into packets for the purpose of multicasting transmissions in the delivery phase.
\begin{figure}
\begin{center}
\vspace{-0.4cm}
\includegraphics[width=0.35\textwidth]{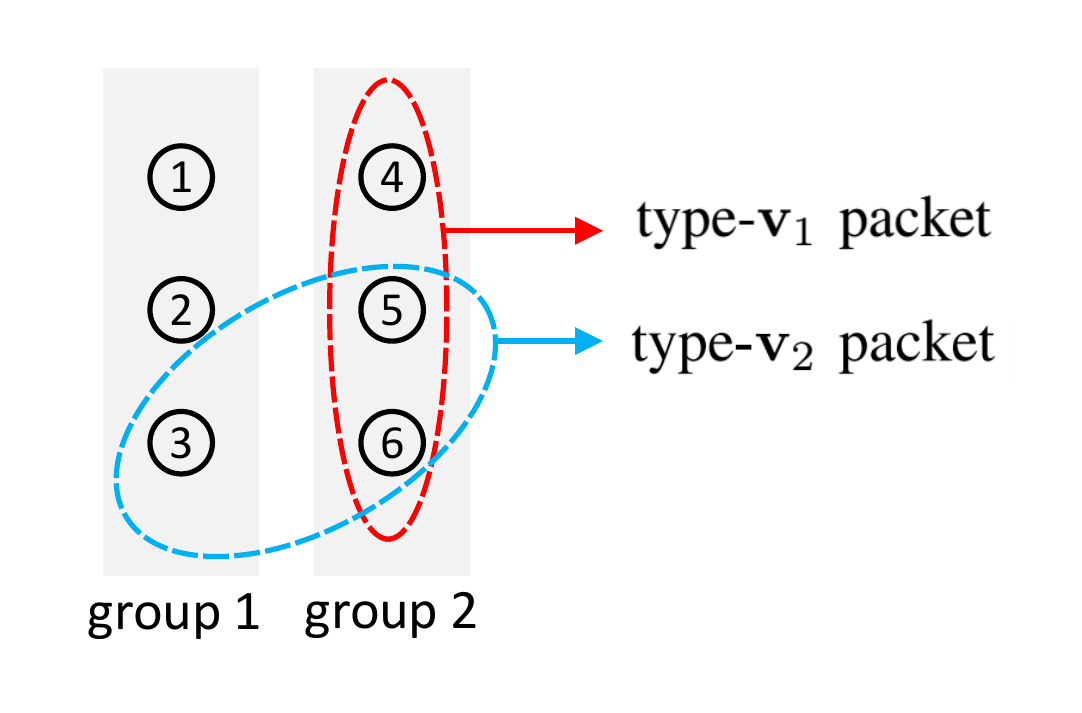}
\vspace{-0.5cm}
\caption{An illustration of packet types under node grouping $\mathbf{q}=(3,3)$.}
\label{figure_illu_pkt_types}
\end{center}
\vspace{-0.5cm}
\end{figure}
\hfill $\lozenge$
\end{example}
}
\fi

\subsubsection{Multicasting Group Type} A \emph{multicasting group} is a set of $t+1$ nodes among which each node broadcasts some packets needed by the remaining $t$ nodes. A \emph{multicasting group type} refers to a specific partition of $t+1$ nodes which is represented by a partition vector $\mathbf{s}:=(s_1,s_2,\cdots,s_{t+1})$ satisfying $\sum_{i=1}^{t+1}s_i=t+1$ and $s_1\geq s_2\geq \cdots s_{t+1}\geq 0$. Different partitions of $t+1$ nodes correspond to different multicasting group types. A \emph{unique group} in $\sv$, denoted by $\bar{\mathcal{U}}_i(i\in \bar{N}_{d})$, refers to the union of parts of $\mathbf{s}$ that contain the same number of nodes, where $\bar{N}_{d}$ denotes the number of distinct parts in $\mathbf{s}$. For a specific multicasting group $\mathcal{S}$ of type $\mathbf{s}$, the set of unique groups of $\sv$ are represented by $\{\bar{\Uc}_{i}\}_{i\in[\bar{N}_{d}]}$ and we have $\mathcal{S}=\bigcup_{i\in[\bar{N}_{d}]}\bar{\Uc}_{i}$. We define a \emph{involved packet type set}, denoted by $\bm{\rho}$, corresponding to a specific multicasting group type $\sv$, as the set of packet types that can appear in the transmission process within multicasting groups of type $\sv$.

\if{0}
We illustrate the above concepts using Example \ref{example_illu_pkt_type} where $K=6$ nodes are partitioned into two groups $\Qc_1=\{1,2,3\}$ and $\Qc_2=\{4,5,6\}$. Note that a multicasting group contains $t+1$ nodes. There are two different multicasting group types, {\em i.e.}, $\sv_1=(3,1)$ and $\sv_2=(2,2)$. A type-$\sv_1$ multicasting group is composed of three nodes from one group and one node from the other group. A type-$\sv_2$ multicasting group is composed of two nodes from both $\Qc_1$ and $\Qc_2$. For example, the multicasting group $\Sc_1$=\{3,4,5,6\} is a type-$\sv_1$ multicasting group which contains one type-$\vv_1$ packet $W_{d_3,\{4,5,6\}}$ and three type-$\vv_2$ packets $W_{d_4,\{3,5,6\}},W_{d_5,\{3,4,6\}}$ and $W_{d_6,\{3,4,5\}}$. Hence, the involved packet type set associated with $\sv_1$ is $\bm{\rho}_1=\{\vv_1,\vv_2\}$. There are $\bar{N}_d=2$ unique groups in $\Sc_1$ which are $\bar{\Uc}_1= \{4,5,6\}$ and $\bar{\Uc}_2= \{3\}$. It can be seen that there are in total $2\binom{3}{3}\binom{3}{1} = 6$ different multicasting groups of type-$\sv_1$. Moreover, $\Sc_2=\{1,2,4,5\}$ is a type-$\sv_2$ multicasting group containing four type-$\vv_2$ packets $W_{d_1,\{2,4,5\}},W_{d_2,\{1,4,5\}},W_{d_4,\{1,2,5\}}$ and $W_{d_5,\{1,2,4\}}$. Since type-$\vv_1$ packets do not appear in $\Sc_2$, the involved packet type set is $\bm{\rho}_2=\{\vv_2\}$. There is only $\bar{N}_d=1$ unique group in $\Sc_2$ which is $\bar{\Uc}_1=\Sc_2$. It is easy to see that there are $\binom{3}{2}^2= 9$ different multicasting groups of type-$\sv_2$.
 \fi

\subsubsection{Further Splitting Ratio (FSR)} The \emph{further splitting ratio} of a packet type $\vv$, denoted by $\alpha(\vv)$, implies that all the type-$\vv$ raw packets need to be split into $\alpha(\vv)\in\mathbbm{Z}^+$ packets in the PTB design. For a multicasiting group $\mathcal{S}$ of type $\mathbf{s}$ containing $\bar{N}_d$ different unique groups, a set of nodes $\mathcal{T}_{\rm x}\subseteq \Sc$ is selected to serve as transmitters for the coded multicasting transmissions in $\Sc$. We can select $\mathcal{T}_{\rm x}$ in such a way that it can be expressed as a union of $|\mathcal{D}_{\rm T}|$ different unique groups {where, in the multicasting group $\Sc$ of type $\sv$, $\mathcal{D}_{\rm T}\subseteq [\bar{N}_d]$ is defined as the set of the indices of the unique groups {\em i.e.}, $\mathcal{T}_{\rm x}=\bigcup_{i\in\mathcal{D}_{\rm T}}\bar{\Uc}_{i}$. Denote $g_i:=|\bar{\Uc}_i|$ as the number of nodes contained in the unique group $\bar{\Uc}_i$, then we have $|\mathcal{T}_{\rm x}|=\sum_{i\in\mathcal{D}_{\rm T}}g_i$. The involved packet type set associated with $\Sc$ contains $\bar{N}_d$ different packet types, {\em i.e.}, $\bm{\rho}=\{\vv_i\}_{i\in[\bar{N}_d]}$. The packet type $\vv_i(i\in[\bar{N}_d])$ is composed of a set of raw packets $\{W_{d_{k_i}, \Sc\backslash\{k_i\}}\}_{k_i\in \bar{\Uc}_i}$. Under such a selection of transmitters, the further splitting ratios for the involved packet types are
\vspace{-0.2cm}
\begin{eqnarray}
\label{equation-splitting_ratio}
\alpha(\mathbf{v}_i)=\left\{\begin{array}{ll}
\sum\limits_{j\in\mathcal{D}_{\rm T}}g_j-1& \text{if}\;i\in\mathcal{D}_{\rm T}\\
\sum\limits_{j\in\mathcal{D}_{\rm T}}g_j &\text{if} \; i\notin\mathcal{D}_{\rm T}
\end{array}\right.
\vspace{-0.2cm}
\end{eqnarray}
which means that each type-$\vv_i$ raw packet needs to be split into $\alpha(\vv_i)$ packets when considering the coded multicasting transmission within the multicasting group $\Sc$ of type $\mathbf{s}$ in the delivery phase. Since one packet type can possibly be contained in multiple involved packet type sets and the above further splitting ratios are derived when only one multicasting group type is considered, we refer to this further splitting ratio as \emph{local further splitting ratio}. 
\if{0}
, which is illustrated in the following using Example \ref{example_illu_pkt_type}.
{
Consider multicasting group $\mathcal{S}_1=\{3,4,5,6\}$ which is composed of two unique groups $\bar{\Uc}_1=\{4,5,6\}$ and $\bar{\Uc}_2=\{3\}$. Each $\Dc_{\rm T}\subseteq [2],\Dc_{\rm T}\neq \emptyset$ corresponds to a specific choice of transmitter selections. Hence, we have three different choices which are $\Dc_{\rm T}=\{1\},\{2\}$ and $\{1,2\}$ respectively. the differences among the three choices are described as follows.

{{\bf \em{Choice 1}}: $\Dc_{\rm T} =\{1\}$. This implies that the first unique group (nodes 4, 5 and 6) is selected as transmitters, \emph{i.e.}, $\Tc_{\rm x} = \bar{\Uc}_1=\{4,5,6\}$, then node 3 will not transmit anything in the delivery phase. In this case, the type-$\vv_1$ raw packet $W_{d_3,\{4,5,6\}}$ needed by node 3 is transmitted by three nodes 4, 5 and 6. To avoid asymmetry in mulicasting transmission and preserve the optimal rate, the packet $W_{d_3,\{4,5,6\}}$ needs to be further split into $\alpha(\vv_1) =3$ packets each of which is delivered by a node in $\bar{\Uc}_1=\{4,5,6\}$. On the other hand, the type-$\vv_2$ raw packet $W_{d_4,\{3,5,6\}}$ needed by node 4 is transmitted by node 5 and 6 (node 3 does not transmit). To preserve symmetry, this packet needs to be further split into $\alpha(\vv_2) = 2$ packets each of which is delivered by node 5 or 6. Similarly, the other two type-$\vv_2$ raw packets $W_{d_4,\{3,4,6\}},W_{d_6,\{3,4,5\}}$ also need to be split into two packets. Since all the raw packets in the JCM scheme are further split into $\alpha^{\rm JCM} :=t$ packets and here we have $\alpha(\vv_2)=2<\alpha^{\rm JCM}=3$, less number of further subpacketization is required, which corresponds to the further splitting ratio gain mentioned previously. We can see that both the raw packet saving gain and the further splitting ratio gain can obtained in this case.}

{{\bf \em{Choice 2}}: $\Dc_{\rm T} =\{2\}$. This implies that the second unique group (node 3) is selected as the only transmitter, \emph{i.e.}, $\Tc_{\rm x} = \bar{\Uc}_2=\{3\}$, then nodes 4, 5 and 6 will only receive packets in the delivery phase. In this case, the type-$\vv_2$ raw packet $W_{d_4,\{3,5,6\}}$ needed by node 4 is only transmitted by node 3. Hence, these is no need to further split it into multiple packets. Similarly, there is no need to split the other two type-$\vv_2$ raw packets $W_{d_5,\{3,4,6\}}, W_{d_6,\{3,4,5\}}$. We use $\alpha(\vv_2)=1$ to indicate that no further splitting is needed. However, the type-$\vv_1$ raw packet $W_{d_3,\{4,5,6\}}$ needed by node 3 is not transmitted by any other node. We can simply exclude this raw packet without sacrificing the completeness of the caching scheme because node 3's need of $W_{d_3,\{4,5,6\}}$ is eliminated though no nodes are transmitting this packet. Actually, all type-$\vv_1$ raw packets can be excluded here, which corresponds to the raw packet saving gain mentioned previously.}

{{\bf \em{Choice 3}}: $\Dc_{\rm T} =\{1,2\} (\emph{JCM scheme})$. This implies that both the two unique groups, \emph{i.e.}, all the four nodes in $\Sc_1$ are selected as transmitters. In this case, any raw packet $W_{d_i,\Sc_1\backslash \{i\}},i\in \Sc_1$, either type $\vv_1$ or type-$\vv_2$, needed by node $i$ is transmitted by the other 3 nodes in $\Sc_1$. To preserve symmetry, all these four packets should be further split into $\alpha(\vv_1) =\alpha(\vv_2) = 3$ packets. We can see that neither the raw packet saving gain nor the further splitting ratio  gain are obtainable here. This actually corresponds to the design of the JCM scheme: For any multicasting group $\Sc$, all the $t+1$ nodes in $\Sc$ are selected as transmitters and a further splitting ratio of $\alpha(\vv) =|\Sc|-1 =t $ is required for any packet type $\vv$, which leads to a subpacketization level of $F^{\rm JCM} = t\binom{K}{t}$. However, as discussed above, it is not always necessary to select all $t+1$ nodes within $\mathcal{S}$ to serve as transmitters. A selection of one or more (not all) unique groups within $\mathcal{S}$ as transmitters will lead to smaller further splitting ratios of the raw packets as indicated in Eq. (\ref{equation-splitting_ratio}), providing an opportunity to reduce the subpacketization, which is the so-called further splitting ratio gain. The overall subpacketization reduction of the proposed design framework is the result of both the raw packet saving gain and the further splitting ratio gain.  }
\fi

\subsubsection{Further Splitting Ratio Table (FSRT)} Given a node grouping $\qv$, denote $V,S$ as the total number of different valid packets types and multicasting types respectively. A \emph{further splitting ratio table} is a matrix $\mathbf{\Lambda }=[\alpha_{ij}]_{S\times V}$ which specifies the local further splitting ratios of packet types derived from all the $S$ multicasting types. More specifically, the $i$-th ($i\in[S]$) row of the FSRT, which is referred to as the \emph{local further splitting ratio vector $\bm{\alpha}_i$}, consists further splitting ratios $\alpha(\mathbf{v}_j)$ for all packet types $\vv_j\in\bm{\rho}_i$ ($\bm{\rho}_i$ is the involved packet set corresponding to multicasting type $\sv_i$) and is specified by Eq. (\ref{equation-splitting_ratio}). All the other entries $\{\alpha(\vv_j):\vv_j\notin \bm{\rho}_i\}$ are left empty. Note that a further splitting ratio of $\alpha=0$ is not the same as an empty entry. To determine the overall further splitting ratio for all the $V$ types of packets, we need to derive the \emph{Least Common Multiple (LCM) vector} $\bm{\alpha}^{\rm LCM}$ (defined below) of the $S$ different local further splitting ratio vectors.

\begin{defn} ({\bf Least Common Multiple (LCM) Vector}) For a set of $n$ vectors $\mathcal{A}=\{\bm{a}_i\}_{i\in[n]}$ in which $|\bm{a}_i|=V$ and $\bm{a}_i$ may contain `empty' entries, the LCM vector of $\mathcal{A}$, denoted by $\bm{a}^{\rm LCM}:={\rm LCM}(\mathcal{A})$, is defined as: $\exists \,z_1,z_2,\cdots,z_n\in\mathbb{Z}^+$ such that: (1) $z_1\bm{a}_1=z_2\bm{a}_2=\cdots= z_n\bm{a}_n$
and (2) $\bm{a}^{\rm LCM}=\arg\min_{z_1\sim z_n}||\bm{a}||_2^2=\arg\min_{z_1\sim z_n}||{ combine}\left(\{z_i\bm{a}_i\}_{i\in[n]}\right)||_2^2$ in which the ${combine}$ operation means that the $j$-th entry of $\bm{a}^{\rm LCM}$ takes the value of the non-zero and non-empty value among the $j$-th entries all the $n$ vectors $\{z_i\bm{a}_i\}_{i\in[n]}$. We assume that 1) the product of any integer and an empty entry is still an empty entry; 2) entry `0' is equal to any other entries, including non-zero entries and empty entries; 3) empty entry is equal to any other zero/non-zero entries. \hfill $\triangle$
\end{defn}

Note that the LCM vector may not always exist. If it exists, it must be unique. In a specific PTB design, the overall splitting ratio vector, denoted by $\bm{\alpha}^{\rm LCM}$, is obtained via deriving the LCM vector of the set of local splitting ratio vectors $\{\bm{\alpha}_i\}_{i\in[S]}$, {\em i.e.}, $\bm{\alpha}^{\rm LCM}:={\rm LCM}\left(\{\bm{\alpha}_i\}_{i\in[S]}\right)$.

\if{0}
{
Still consider
Example~\ref{example_illu_pkt_type}. 
 There are $S=2$ multicasting group types $\mathbf{s}_1=(3,1)$ and $\mathbf{s}_2=(2,2)$. There are also $V=2$ packets types $\mathbf{v}_1=(3,0)$ and $\mathbf{v}_2=(2,1)$. The corresponding involved packet type sets are $\bm{\rho}_1=\{\mathbf{v}_1,\mathbf{v}_2\}$ and $\bm{\rho}_2=\{\mathbf{v}_2\}$. The selection of transmitters is as follows. For $\mathbf{s}_1$, we choose the second unique group as transmitters. We use the superscript $^{\star}$ to mark the transmitters within a multicasting group type, {\em i.e.}, $\mathbf{s}_1=(3,1^{\star})$. For example, in a specific multicasting group $\mathcal{S}_1=\{3,4,5,6\}$, the transmitters is node 3. This selection will result in a local further splitting ratio vector $\bm{\alpha}_1:=\left(\alpha(\mathbf{v}_1),\alpha(\mathbf{v}_2) \right)=(0,1)$. For $\mathbf{s}_2$, the only choice is to select all $t+1=4$ nodes as transmitters since there is only $N_{d}=1$ unique group which has to be selected. This results in a further splitting ratio vector $\bm{\alpha}_2:=(\bullet,\alpha(\mathbf{v}_2))=(\bullet,3)$ in which the symbol $\bullet$ denotes an empty entry since type-$\mathbf{v}_1$ packets do not appear in type-$\mathbf{s}_2$ multicasting groups. As a result, the FSRT is
\begin{center}
\begin{tabular}{|c| c c|}
\hline
 &  $\mathbf{v}_1$ & $\mathbf{v}_2$  \\
\hline\hline
$\bm{\alpha}_1$ & 0  &1  \\
\hline
$\bm{\alpha}_2$  &$\bullet$ &3 \\
 \hline
\end{tabular}
\end{center}
from which we can easily obtain $\bm{\alpha}^{\rm LCM}={\rm LCM}(\bm{\alpha}_1,\bm{\alpha}_2)=(0,3)$, implying that in the PTB design, type-$\mathbf{v}_1$ raw packets are excluded while each type-$\mathbf{v}_2$ raw packet is further split into 3 packets.
}
\fi

\subsubsection{Memory Constraint Table (MCT)} Given a node grouping $\qv$ containing $N_d$ different unique groups, a \emph{memory constraint table} is a matrix $\mathbf{\Omega}=[\omega_{ij}]_{N_{d}\times V}$ with $\omega_{ij}:=F_i(\mathbf{v}_j)$ where $F_i(\mathbf{v}_j)$ denotes the number of raw packets of type $\vv_j$ cached by a node in the $i$-th unique group. Denote  $\mathbf{F}_i:=[F_i(\vv_1),F_i(\vv_2),\cdots, F_i(\vv_V)]$ as the $i$-th row of  $\mathbf{\Omega}$. Also denote the \emph{raw packet number vector} as $\mathbf{F}:=[F(\vv_1),F(\vv_2),\cdots, F(\vv_V)]$ where $F(\vv_j)$ represents the number of raw packets of type $\vv_j(j\in[V])$ in a PTB design.
Furthermore, $\forall i\in [N_{\rm d}-1]$, we define the \emph{node cache difference vector} as $\Delta\mathbf{F}_{i}:=(f_{i1},f_{i2},\cdots,f_{iV})=\mathbf{F}_{i+1}-\mathbf{F}_{i}$ in which $f_{ij}=F_{i+1}(\mathbf{v}_j)-F_{i}(\mathbf{v}_j),\forall j\in[V]$ is the difference of the number of type-$\mathbf{v}_j$ raw packets cached by nodes in the $(i+1)$-th and $i$-th unique group $\mathcal{U}_{i+1}$ and $\mathcal{U}_{i}$. Let all the packets have the same size, the memory constraint can be represented as $\bm{\alpha}^{\rm LCM}{\Delta\mathbf{F}_i}^{\rm T}=\bm{\alpha}^{\rm LCM}({\mathbf{F}_{i+1}}-{\mathbf{F}_i})^{\rm T}=0,\forall i\in[1:N_{d}-1]$, {\em i.e.}, $\bm{\alpha}^{\rm LCM}{\mathbf{F}_{1}}^{\rm T}=\bm{\alpha}^{\rm LCM}{\mathbf{F}_2}^{\rm T}
\cdots=\bm{\alpha}^{\rm LCM}{\mathbf{F}_{N_{d}}}^{\rm T}$, implying that nodes in all the $N_{d}$ unique groups have cached the same number of packets. Since all the nodes have identical cache memory size, caching the same number of packets of equal length satisfies the memory constraint. The exact length of the packets can be determined by the fact that each node has a cache memory size of $M$ files.

\if{0}
Consider Example~\ref{example_illu_pkt_type}. Since the node grouping $\mathbf{q}=(3,3)$ is an equal grouping and there is only one unique group ($N_d=1$), the node caching memory constraint is automatically satisfied. The number of raw packets is equal to $\mathbf{F}:=\left(F(\vv_1),F(\vv_2)\right)=(2,18)$ and the overall further splitting ratio vector is $\bm{\alpha}:=(\alpha(\vv_1), \alpha(\vv_2))=(0,3)$. Hence, the total number of packets  required is equal to $F=\bm{\alpha}^{\rm LCM}\mathbf{F}^{\rm T}=(0,3)(2,18)^{\rm T}=54$, which is less than $F^{\rm JCM}=3\binom{6}{3}=60$. Note that the subpacketization reduction comes from excluding two type-$\mathbf{v}_1$ raw packets. Hence, only raw packet saving gain is available in this example.
\fi

\subsubsection{PTB Design as An Integer Optimization} With all the above definitions, under the condition of equal-length subpacketizations (all packets have identical length), the integer optimization problem that determines the optimal LCM vector which results in the minimum $F$ is given by Eq. (\ref{eq: optimization 1}) to (\ref{eq: optimization 3}), where $\Phi$ represents the set of all possible LCM vectors derived from the $S$ local further splitting vectors based on the set of all possible node grouping $\mathbf{q}$ and the set of all possible selections of transmitters within each multicasting group type under each $\mathbf{q}$.
Although each feasible solution of the above optimization problem corresponds to a valid PTB design which may or may not yield a lower subpacketization level than the JCM scheme,
in this paper, we present several PTB designs with order or constant reduction on the subpacketization levels compared to the JCM scheme, implying that the JCM scheme is far from optimal in terms of subpacketization. The extension of the optimization problem (\ref{eq: optimization 1}) to (\ref{eq: optimization 3}) to the case of unequal-length subpacketizations (different packets may have distinct packet length) can be found in \cite{zhang2018D2Dlong}. 

\if{0}
Moreover, we can extend the optimization problem (\ref{eq: optimization 1}) to (\ref{eq: optimization 3}) to the case of unequal-length subpacketizations (different packets may have distinct packet length) as follows.
\vspace{-0.1cm}
\begin{eqnarray}
&\min& F:=\left(\sum_{h=1}^{H}\bm{\alpha}^{(h)}\right)\mathbf{F}^{\rm T} \label{eq: optimization 4} \\
&{\rm s.t.}&\bm{\alpha}^{(h)}\in \Phi,\;\forall h\in[H], \label{eq: optimization 5} \\
&\,&\sum_{h=1}^{H}\gamma_h\bm{\alpha}^{(h)}\left(\Delta\mathbf{F}_i^{(h)}\right)^{\rm T}=0,\;\forall i\in[N_{\rm d}-1], \label{eq: optimization 6}
\vspace{-0.2cm}
\end{eqnarray}
where $H$ denotes the number of different \emph{coupled groups} $\mathcal{G}_h,h\in[H]$, which is defined as a set of packet types with the same packet size $\ell_h$ bits. 
$\gamma_h$ denotes the packet size ratio of $\mathcal{G}_h$ over $\mathcal{G}_1$, {\em i.e.}, $\gamma_h:=\frac{\ell_h}{\ell_1}$. Moreover, $\bm{\alpha}^{(h)}$ denotes the 
LCM vector derived based on the coupled group $\mathcal{G}_h$ and $\Delta\mathbf{F}_i^{(h)}$ represents the node cache difference vector based on the coupled group $\mathcal{G}_h$.
\fi

\if{0}
\vspace{-0.1cm}
\section{An Example}
\label{sec: example}

In this section, we illustrate the above combinatorial design framework via a concrete and complete example. The caching parameter is assumed to be integers, \emph{i.e.}, $t:=\frac{KM}{N}\in\mathbbm{Z}^+$. We also denote $\bar{t}:=K-t=K\left(1-\frac{M}{N}\right)$.

Now consider an example with parameters $(K,\bar{t})=(3m,3)$ where $m\geq \bar{t}=3$ and node grouping $\mathbf{q}=(3,3,\cdots,3)$. In this case, $(S,V)=(2,3)$. The packet types, multicasting group types and involved packet type sets are (the unique groups which are selected as transmitters within each multicasting group type are marked with the superscript $^\star$) :
\vspace{-0.1cm}
\begin{eqnarray}
\mathbf{v}_1&=&\big(\mathbf{3}^{(m-3)},\mathbf{2}^{(3)}\big), 
\mathbf{v}_2 = \big(\mathbf{3}^{(m-2)},\mathbf{2}^{(1)},\mathbf{1}^{(1)}\big)\nonumber\\
\mathbf{v}_3&=&\big(\mathbf{3}^{(m-1)},\mathbf{0}^{(1)}\big),  
\mathbf{s}_1= \big(\mathbf{3}^{(m-1)},\mathbf{1}^{(1)\star}\big),\;\bm{\rho}_1=\{\mathbf{v}_2,\mathbf{v}_3\}\nonumber\\
\mathbf{s}_2&=&\big(\mathbf{3}^{(m-2)},\mathbf{2}^{(2)\star}\big),\;\bm{\rho}_2=\{\mathbf{v}_1,\mathbf{v}_2\}\nonumber\nonumber
\end{eqnarray}
\vspace{-0.1cm}
and the further splitting ratio table is
\begin{center}
\begin{tabular}{|c| c c c|}
\hline
 &  $\mathbf{v}_1$ & $\mathbf{v}_2$ & $\mathbf{v}_3$  \\
\hline\hline
$\bm{\alpha}_1$ &  $\bullet$ &1 &0  \\
\hline
$\bm{\alpha}_2$ &4 & 3 & $\bullet$\\
 \hline
 $\bm{\alpha}^{\rm LCM}$ &4 & 3 &0\\
 \hline
\end{tabular}
\end{center}
from which we obtain $\bm{\alpha}^{\rm LCM}=(4,3,0)$, implying that $\mathbf{v}_3$ is excluded (\emph{raw packet saving gain}) and $\mathbf{v}_1,\mathbf{v}_2$ have further splitting ratios of $4,3$ respectively (\emph{further splitting ratio gain}). Hence, the number of packets is equal to
\be
F=(4,3,0)[F(\mathbf{v}_1),F(\mathbf{v}_2),F(\mathbf{v}_3)]^{\rm T} 
=\frac{K(K-3)(2K-3)}{3},
\ee
where $F(\mathbf{v}_1)=\binom{m}{3}\binom{3}{2}^3=\frac{K(K-3)(K-6)}{6}$, $F(\mathbf{v}_2)=\binom{m}{1}\binom{3}{2}\binom{m-1}{1}\binom{3}{1}=K(K-3)$ and $F(\mathbf{v}_3)=\binom{m}{1}\binom{3}{3}=\frac{K}{3}$.
It can be seen that when $K \geq 9$,
$F/F^{\rm JCM} 
= 4/K + o\left(4/K\right)$,
which means an order gain of packetization level is obtained using the proposed scheme compared to the JCM scheme.


{
Next we consider the detailed delivery procedure for a system with $K=9,N=3,M=2$ and $t=6$. Consider a specific equal-grouping assignment $\mathcal{Q}_1=\{1,2,3\},\mathcal{Q}_2=\{4,5,6\}$ and $\mathcal{Q}_3=\{7,8,9\}$.
Since $\bm{\alpha}^{\rm LCM}=(4,3,0)$, type-$\mathbf{v}_3$ packets are excluded and type-$\mathbf{v}_1$ and type-$\mathbf{v}_2$  packets need to be further split into 4 and 3 packets respectively. In this case $F=270$ while $F^{\rm JCM}=504$. The cache placement is that node $k$ stores any packet $W_{n,\mathcal{T}}^{(j)}$ if $k\in\mathcal{T}$. Note that in this example the subpacketization reduction gain compared to the JCM scheme consists of two parts: \emph{1) raw packet saving gain}: $tF(\vv_3)=18$ packets and \emph{2) further splitting ratio gain}: $(t-4)F(\vv_1)+(t-3)F(\vv_2)=216$  packets. We can see that the reduction is mainly due to smaller further splitting ratios of type-$\vv_1$ and type-$\vv_2$ packets.

For a type-$\mathbf{s}_1$ multicasting group $\mathcal{S}_1=[7]$, node 7 is the only transmitter and it transmits three coded multicast messages $\bigoplus_{k\in[6]}W_{d_k,\mathcal{S}_1\backslash \{k\}}^{(j)}$, $j=1,2,3$ to other nodes in $\mathcal{S}_1$. Each node $k$ recovers its desired packets $\{W_{d_k,\mathcal{S}_1\backslash \{k\}}^{(j)}:j=1,2,3\}$ with the help of the cached content while node 7 itself only transmits but receives nothing. For a type-$\mathbf{s}_2$  multicasting group $\mathcal{S}_2=[9]\backslash\{6,9\}$, the set of type-$\vv_1$ and $\vv_2$ packets involved are $\{W_{d_k,\mathcal{S}_2\backslash\{k\}}^{(j)}:j\in[4],k\in[3]\}$ and $\{W_{d_k,\mathcal{S}_2\backslash\{k\}}^{(j)}:j\in[3],k\in \mathcal{S}_2\backslash\mathcal{Q}_1\}$ respectively. Denote $W^{(j)}:=\bigoplus_{k\in[3]}W_{d_k,\mathcal{S}_2\backslash\{k\}}^{(j)},j\in[4]$. Nodes 4, 5, 7, 8 each sends a coded multicast message as follows.
\begin{eqnarray}
W_4&=&W^{(1)}\oplus W_{d_5,\mathcal{S}_2\backslash\{5\}}^{(1)}\oplus W_{d_7,\mathcal{S}_2\backslash\{7\}}^{(1)} \oplus W_{d_8,\mathcal{S}_2\backslash\{8\}}^{(1)}\nonumber\\
W_5&=&W^{(2)}\oplus W_{d_4,\mathcal{S}_2\backslash\{4\}}^{(1)}\oplus W_{d_7,\mathcal{S}_2\backslash\{7\}}^{(2)} \oplus W_{d_8,\mathcal{S}_2\backslash\{8\}}^{(2)}\nonumber\\
W_7&=&W^{(3)}\oplus W_{d_4,\mathcal{S}_2\backslash\{4\}}^{(2)} \oplus W_{d_5,\mathcal{S}_2\backslash\{5\}}^{(2)} \oplus W_{d_8,\mathcal{S}_2\backslash\{8\}}^{(3)}\nonumber\\
W_8&=&W^{(4)}\oplus  W_{d_4,\mathcal{S}_2\backslash\{4\}}^{(3)}\oplus W_{d_5,\mathcal{S}_2\backslash\{5\}}^{(3)}\oplus W_{d_7,\mathcal{S}_2\backslash\{7\}}^{(3)} \nonumber
\end{eqnarray}
from which we can see that all nodes can recover their desired packets. Since each coded message is simultaneously useful for $t=6$ nodes, the rate is optimal. The transmission procedure for other multicasting groups is similar.
}
\fi

\if{0}
\section{Main Results and Discussions}
\label{section: main result}

\begin{theorem}
\label{theorem_1_large_t}
For even $\bar{t}:=K-t$, where  
$K=2m$, using the PTB design framework, the optimal rate of D2D caching networks is achievable and  
\begin{eqnarray}
\frac{F}{F^{\rm JCM}}=\Theta\left(\frac{f(\bar{t})}{K-\bar{t}}\right),
\end{eqnarray}
where $f(\bar{t}):=\prod_{i=1}^{\frac{\bar{t}}{2}}(2i-1)$ is a function which depends only on $\bar{t}$. Moreover, $\forall K \geq 2\bar{t}$, and $\bar t = O(\log\log K)$, 
$\frac{F}{F^{\rm JCM}}$ 
vanishes as $K$ goes to infinity. 
\hfill $\square$
\end{theorem}

From Theorem \ref{theorem_1_large_t}, it can be seen that when $K \leq 2  t$ is even and $t$ is large enough ({\em i.e.}, $t = K - O(\log\log K)$), an order gain in terms of subpacketization can be obtained using the PTB design compared to the JCM scheme 
while preserving the optimal rate. However, it can be seen that for small $t$, the PTB design achieving Theorem \ref{theorem_1_large_t} may result in an even worse subpacketization compared to the JCM scheme. In the following theorem, we provide a general result for even $K$ and $t$ based on a specific PTB design and provide subpacketization gains compared to the JCM scheme when $t$ is small.

\begin{theorem}
\label{theorem_2_groups}
For $(K,t)=(2q,2r)$ with $q\geq t+1$ and $r\geq 1(r\in \mathbbm{Z}^+)$, using PTB design framework with the two-group equal grouping, {\em i.e.}, $\mathbf{q}=(\frac{K}{2},\frac{K}{2})$, the optimal rate of D2D caching networks is achievable by the further splitting ratio vector $\bm{\alpha}^{\rm LCM}=(0,1,2,\cdots,r)$. Further, when $r \geq 2$, we have $\frac{F}{F^{\rm JCM }}< \frac{1}{2}\left(1-\frac{1}{2^{t-1}}\right) $ and $\zeta(t)=\frac{1}{2} $.
\hfill $\square$
\end{theorem}

From Theorem \ref{theorem_2_groups}, it can be seen that under the conditions of even $K$ and $t$, based on a 2-group equal grouping PTB design, a constant gain of a multiplicative factor less than $0.5$ can be always achieved compared to the JCM scheme. This design applies to the case of any values of $t$ as long as it is an even number. When $K$ is odd,
it is surprisingly more difficult than the case of even $K$. In Section~\ref{sec: example}, we provided an example showing that when $K=3m$ and $t = K-3$, $m \in \mathbb{Z}^+$, it is possible to exploit an equal user grouping to achieve an order gain of subpacketization level compared to the JCM scheme. However, in general, we may need to use the general PTB design framework in (\ref{eq: optimization 4}) to (\ref{eq: optimization 6}) that exploits the heterogeneous packet size.

\begin{theorem}
\label{theorem_heter_pkt_size}
For $(K,t)=(2q+1,2r)$ with $q\geq 2r+1,r\geq 1 (r\in\mathbbm{Z}^+)$, using the two-group unequal grouping $\mathbf{q}=(\frac{K+1}{2},\frac{K-1}{2})$, the optimal rate of D2D caching networks is achievable by the further splitting ratio vector $\bm{\alpha}^{\rm LCM}=(0,2,4,\cdots,t-2,t,t,\cdots, t)$ and we have $$\frac{F}{F^{\rm JCM}}< \frac{1}{t}\left(\frac{\binom{t}{r}}{2^t}-1\right)+1.$$
\hfill $\square$
\end{theorem}

From Theorem \ref{theorem_heter_pkt_size}, it can be seen that by using the general PTB design framework in (\ref{eq: optimization 4}) to (\ref{eq: optimization 6}) with the consideration of heterogeneous packet size, when $K$ is odd, a constant gain in terms of subpacketization compared to the JCM scheme can be achieved while preserving the optimal rate when $t$ small.
\fi

\if{0}
\section{Achievable Scheme}
\label{section_achievable_scheme}
\subsection{PTB Design for Large $t$}
In this section we present the achievable scheme of Theorem \ref{theorem_1_large_t} and some examples to show that order reduction of subpacketization level can be achieved while preserving the same rate as the JCM scheme. We assume that $K$ is sufficiently large while $\bar{t}:=K-t$ is kept fixed. We will show later that $\bar{t}$ can be extended to higher order.
\label{subsection_large_t}
For $(K,\bar{t})=(2m,2r)$ with $m\geq \bar{t}+1,r\geq 1$, consider the node grouping $\mathbf{q}=(2,2,\cdots,2)$ with $m=\frac{K}{2}$ groups each containing exactly two nodes. In this case, the number of multicasting types and packet types are $(S,V)=(r,r+1)$. More specifically, the $i$-th ($i\in[r]$) multicasting type $\mathbf{s}_i$ and the $j$-th ($j\in[r+1]$) packet type $\mathbf{v}_j$ are (transmitters are marked by the superscript $^\star$)\footnote{For simplicity, we use the notation $\mathbf{s}=\big(\mathbf{a}^{(x)},\mathbf{b}^{(y)},\mathbf{c}^{(z)},\cdots\big)$ to denote that there are $x$ parts equal to $a$, $y$ parts equal to $b$, and $z$ parts equal to $c$ and so on.}
\begin{eqnarray}
\mathbf{s}_i&=&\big(\mathbf{2}^{(m-(r+i)+1)},\mathbf{1}^{(2i-1)\star},\mathbf{0}^{(r-i)}\big)\\
\mathbf{v}_j&=&\big(\mathbf{2}^{(m-(r+j)+1)},\mathbf{1}^{(2(j-1))},\mathbf{0}^{(r-j+1)}\big)
\end{eqnarray}

The $i$-th involved packet type set is $\bm{\rho}_i=\{\mathbf{v}_{i+1},\mathbf{v}_i\}$ and the corresponding local further splitting ratio vector is $\bm{\alpha}_i:=\left[\alpha(\mathbf{v}_{i}),\alpha(\mathbf{v}_{i+1})\right]=\left[2(i-1),2i-1 \right]$.\footnote{Note that we ignored the 'empty' entries for packet types other than $\mathbf{v}_{j}$ and $\mathbf{v}_{j+1}$ for ease of notation.} The corresponding further splitting ratio table is shown in Table \ref  .As a result, the overall LCM vector will be $\bm{\alpha}^{\rm LCM}=(\alpha_1,\alpha_2,\cdots,\alpha_{r+1})$ in which the entries are $\alpha_1=0$ (meaning $\mathbf{v}_1$ is excluded), $\alpha_2=2^{r-1}\prod_{i=1}^{r-1}i,\,\alpha_{r}=2(r-1)\prod_{i=1}^{r-1}(2i-1)$ and $\alpha_{r+1}=\prod_{i=1}^{r}(2i-1)$. For $j\in[3:r-1]$,\footnote{Due to space limit, the detailed derivation of these ratios are omitted.}
\begin{eqnarray}
\alpha_j=\prod_{i=1}^{j-1}(2i-1)\left(\prod_{i=j-1}^{r-1}2i\right)
\end{eqnarray}

Hence, the number of subpackets required is equal to
\begin{eqnarray}
F=\bm{\alpha}^{\rm LCM}\mathbf{F}^{\rm T}=\sum_{j=1}^{r+1}\alpha_j F(\mathbf{v}_j)
\end{eqnarray}
in which $\mathbf{F}:=[F(\mathbf{v}_1),F(\mathbf{v}_2),\cdots,F(\mathbf{v}_{r+1})]$ denotes the number of raw packets corresponding to the $V=r+1$ packet types.

\begin{table*}
\caption{Further Splitting Ratio Table for $m=\frac{K}{2}$ Eqaul-grouping PTB Design}
\begin{center}
\begin{tabular}{|c| c c c c c c c c c  c c |}
\hline
  & $\mathbf{v}_1$ & $\mathbf{v}_2$ & $\mathbf{v}_3$ &$\mathbf{v}_4$&$\cdots $&$\mathbf{v}_j$ &$\mathbf{v}_{j+1}$ &$\mathbf{v}_{j+2}$ &$\cdots $ &$\mathbf{v}_{r}$ & $\mathbf{v}_{r+1}$\\
 \hline\hline
  $\mathbf{s}_1$ &0& 1  &   &  &  & &   & & & & \\
 \hline
   $\mathbf{s}_2$&& 2 & 3  &  &  &  & &  & &&  \\
 \hline
   $\mathbf{s}_3$& &  & 4  & 5 &  &  &  &  && & \\
 \hline
 $\vdots$ & &  &   &   &$\ddots$  &    &&  & && \\
 \hline
  $\mathbf{s}_j$ & &  &   & &  & $2(j-1)$&$2j-1$  & &  &&  \\
 \hline
   $\mathbf{s}_{j+1}$ & &  &   &  &  &  &$2j$ &$2j+1$& & &  \\
 \hline
   $\vdots $ & &  &   &  &  &  &  &  &$\ddots$&&   \\
 \hline
   $\mathbf{s}_r$ & &  &   &  &  &  &  &&&  $2(r-1)$& $2r-1$  \\
 \hline
    $\bm{\alpha}^{\rm LCM}$ &$\alpha_1$ & $\alpha_2$ & $\alpha_3$  & $\alpha_3$  & $\cdots$ &$\alpha_j$ &$\alpha_{j+1}$  &$\alpha_{j+2}$&$\cdots$ & $\alpha_r$ &$\alpha_{r+1}$  \\
    \hline
\end{tabular}
\end{center}\label{table_m_grps_q=2}
\end{table*}

Regarding $\bm{\alpha}^{\rm LCM}$, we have the following lemma, whose proof is omitted due to space limit.
\begin{lemma}
\label{lemma_monotonity_alpha}
\emph{For the derived LCM vector $\bm{\alpha}^{\rm LCM}$, the entry sequence $\{\alpha_j\}_{j=1,2,\cdots,r+1}$ is strictly increasing.}
\end{lemma}

To derive an upper bound on $\frac{F}{F^{\rm JCM}}$, we have the following definition and lemma.
\begin{defn} \emph{(Dominant Packet Type)}
Under a specific node grouping, the dominant packet type, denoted by $\mathbf{v}_{\rm dom}$, is defined as the packet type containing the largest number of raw packets for sufficiently large $K$.  \hfill $\triangle$
\end{defn}
\begin{lemma}
\label{lemma_dominant_type}
\emph{For a specific packet type $\mathbf{v}$, which is a partition of $\bar{t}$,\footnote{Each type of packets can be represent by a specific partition of $t$, or alternatively, it can be represented as a partition of $\bar{t}=K-t$, which means deleting $\bar{t}$ nodes from the node grouping $\mathbf{q}$ and the remaining $K-\bar{t}=t$ nodes form a packet type, i.e, the packets of this type are exclusively cached in the remaining $t$ nodes.} let $N_{\rm d},N_{\rm p}$ denote the number of distinct parts and parts of $\mathbf{v}$ respectively. For a given equal node grouping $\mathbf{q}=(q,q,\cdots,q)$ where $q=\frac{K}{m}\in\mathbb{Z}^+$ and $m\geq t+1$, the dominant packet type is the one with $N_{\rm p}=\bar{t}$, {\em i.e.}, $\mathbf{v}_{\rm dom}=(q-1,q-1,\cdots,q-1)$. }
\end{lemma}

\emph{Proof:} Recall that a packet type $\mathbf{v}$ can be equivalently represented as a sequence $\{(\beta_i,\psi_i)\}_{i=1,2,\cdots,N_{\rm d}}$ satisfying $\beta_1\geq \beta_2\geq \cdots \geq \beta_{N_{\rm d}}>0$ and $\sum_{i=1}^{N_{\rm d}}\beta_i\psi_i=\bar{t},\sum_{i=1}^{N_{\rm d}}\psi_i=N_{\rm p}$. Then the number of raw packets corresponding to each type can be calculated as (assuming $N_{\rm d}\geq 2$)
\begin{eqnarray}
\label{number_of_pkts}
F(\mathbf{v})&=&\prod_{i=1}^{N_{\rm d}}\binom{m-d_i}{\psi_i}\binom{q}{\beta_i}^{\psi_i}\nonumber\\
&=&\prod_{i=1}^{N_{\rm d}}\prod_{j=0}^{\psi_i-1}(m-d_i-j)f(\beta_i,\psi_i)\nonumber\\
&\overset{(a)}{\approx}&\prod_{i=1}^{N_{\rm d}}m^{\psi_i}f(\beta_i,\psi_i)\nonumber\\
&=&m^{\sum_{i=1}^{N_{\rm d}}\psi_i}\prod_{i=1}^{N_{\rm d}}f(\beta_i,\psi_i)\nonumber\\
&=&m^{N_{\rm p}}\prod_{i=1}^{N_{\rm d}}f(\beta_i,\psi_i)\nonumber\\
&=&K^{N_{\rm p}}\frac{\prod_{i=1}^{N_{\rm d}}f(\beta_i,\psi_i)}{q^{N_{\rm p}}}\nonumber\\
&=&CK^{N_{\rm p}}
\end{eqnarray}
where in $(a)$ we let $K$ be sufficiently large and keep $q$ fixed. The notations are as follows: $\psi_0:=0$ and $d_i:=\sum_{j=1}^{i}\psi_{j-1}$. The function $f$ is defined as $f(\beta_i,\psi_i):=\binom{q}{\beta_i}^{\psi_i}/\psi_i!$ and $C=\frac{\prod_{i=1}^{N_{\rm d}}f(\beta_i,\psi_i)}{q^{N_{\rm p}}}$ is a constant which does not depend on $K$.

From (\ref{number_of_pkts}) we see that $F(\mathbf{v})=\Theta\left(K^{N_{\rm p}}\right)$. Hence, the dominant type will be the type with maximum $N_{\rm p}=t$, {\em i.e.}, $\mathbf{v}_{\rm dom}=\mathbf{v}_{r+1}=(q-1,q-1,\cdots,q-1)$. \hfill $\square$

Since $F(\mathbf{v}_{\rm dom})=\Theta\left(K^{\bar{t}}\right)$ is at least an order greater than other packet types, for large $K$, we can approximate $F$ by $F(\mathbf{v}_{\rm dom})$, {\em i.e.},
\begin{eqnarray}
F\approx \alpha_{r+1} F(\mathbf{v}_{r+1})&=& \alpha_{r+1}\binom{m}{\bar{t}}q^{\bar{t}}\nonumber\\
&=& \alpha_{r+1}\frac{1}{\bar{t}!}\prod_{i=0}^{\bar{t}-1}\left(\frac{K}{q}-i\right)q^{\bar{t}}\nonumber\\
&=& \alpha_{r+1}\frac{1}{\bar{t}!}\prod_{i=0}^{\bar{t}-1}\left(K-iq\right)\nonumber\\
&\approx& \alpha_{r+1}\frac{K^{\bar{t}}}{\bar{t}!}
\end{eqnarray}
where $ \alpha_{r+1}=\prod_{i=1}^{r}(2i-1)$. Note that $F^{\rm JCM}\approx(K-\bar{t})\frac{K^{\bar{t}}}{\bar{t}!}$ for large $K$. Hence, for sufficiently large $K$,
\begin{eqnarray}
\frac{F}{F^{\rm JCM}}&\approx&\frac{\prod_{i=1}^{\frac{\bar{t}}{2}}(2i-1)}{K-\bar{t}}
\end{eqnarray}
or equivalently, $\frac{F}{F^{\rm JCM}}=\Theta\left(\frac{f(\bar{t})}{K-\bar{t}}\right)$ with $f(\bar{t}):=\prod_{i=1}^{\frac{\bar{t}}{2}}(2i-1)$, which vanishes as $K$ grows large, yielding an order reduction on subpacketization.

Next we derive a strict upper bound on the ratio $\frac{F}{F^{\rm JCM}}$. Due to Lemma \ref{lemma_monotonity_alpha}, we have $\alpha_i\leq \alpha_{r+1},\forall j\in[r+1]$. As a result, for any $K\geq 2\bar{t}$, we have
\begin{eqnarray}
\frac{F}{F^{\rm JCM}}&=&\frac{\sum_{j=1}^{r+1}\alpha_j F(\mathbf{v}_j)}{t\sum_{j=1}^{r+1}F(\mathbf{v}_j)}\nonumber\\
&<&\frac{\alpha_{r+1}}{t}\frac{\sum_{j=1}^{r+1} F(\mathbf{v}_j)}{\sum_{j=1}^{r+1}F(\mathbf{v}_j)}\nonumber\\
&=&\frac{\alpha_{r+1}}{K-\bar{t}}\nonumber\\
&=&\frac{f(\bar{t})}{K-\bar{t}}
\end{eqnarray}

The next lemma shows that the upper bound $\frac{f(\bar{t})}{K-\bar{t}}$ vanishes if $\bar{t}\leq \Theta\left(\log\log K\right)$. A special case is that $\bar{t}$ is kept fixed as a constant when $K$ grows.
\begin{lemma}
\label{lemma_vanishing_bound}
\emph{For $\bar{t}:=K-t\leq \Theta\left(\log\log K\right)$, the upper bound $\frac{f(\bar{t})}{K-\bar{t}}=\frac{\prod_{i=1}^{\frac{\bar{t}}{2}}(2i-1)}{K-\bar{t}}$ on $\frac{F}{F^{\rm JCM}}$ vanishes as $K\to\infty$.}
\end{lemma}

\emph{Proof:} WLOG, assume $\bar{t}=2c\log\log K$ for some constant $c$ and $K=2^{2^\kappa}$. Then
\begin{eqnarray}
\lim\limits_{K\to\infty}\frac{F}{F^{\rm JCM}}&=&\lim\limits_{K\to\infty}\frac{\prod_{i=1}^{c\log\log K}(2i-1)}{K-2c\log\log K}\nonumber\\
&=&\lim\limits_{K\to\infty}\frac{\prod_{i=1}^{\kappa c}(2i-1)}{2^{2^{\kappa}}-2\kappa c}\nonumber\\
&=&\lim\limits_{K\to\infty}\frac{\prod_{i=1}^{\kappa c}(2i-1)}{2^{2^\kappa}}\nonumber\\
&=&\lim\limits_{K\to\infty}\prod_{i=1}^{\kappa c}\frac{2i-1}{2^{\frac{2^\kappa}{\kappa c}}}
\end{eqnarray}

Since each individual term $f_i:=\frac{2i-1}{2^{\frac{2^\kappa}{\kappa c}}},i\in[ \kappa c]$ satisfies
\begin{eqnarray}
f_i:=\frac{2i-1}{2^{\frac{2^\kappa}{\kappa c}}}\leq\frac{2\kappa-1}{2^{\frac{2^\kappa}{\kappa c}}}\overset{K\to\infty}{=}0
\end{eqnarray}
implying $\lim\limits_{K\to\infty}f_i=0$. As a result, we obtain
\begin{eqnarray}
\lim\limits_{K\to\infty}\frac{F}{F^{\rm JCM}}=\lim\limits_{K\to\infty}\prod_{i=1}^{\kappa c}f_i=\prod_{i=1}^{\kappa c}\lim\limits_{K\to\infty}f_i=0
\end{eqnarray}
which completes the proof of Theorem \ref{theorem_1_large_t}. \hfill $\square$

\begin{remark}
The node grouping $\mathbf{q}=(2,2,\cdots,2)$ actually achieves the minimum $F$ among a set of equal grouping designs, which is shown in the following lemma.
\end{remark}

\begin{lemma}
\emph{For $(K,\bar{t})=(mq,2)$, the node grouping $\mathbf{q}=(2,2,\cdots,2)$ yields the minimum $F$ among the set of equal groupings $\{\mathbf{q}=(q,q,\cdots,q):q\geq 2,mq=K\}$}.
\end{lemma}

\emph{Proof}: For the PTB design based on the grouping $\mathbf{q}=(q,q,\cdots,q)(q\geq 2)$, we have $(S,V)=(1,2)$. The packet and multicasting group types are (transmitters are marked by the superscript $^{\star}$)
\begin{eqnarray}
\mathbf{v}_1&=&\big(q^{(m-1)},q-2\big)\nonumber\\
\mathbf{v}_2&=&\big(q^{(m-2)},(q-1)^{(2)}\big)\nonumber\\
\mathbf{s}&=&\big(q^{(m-1)},(q-1)^{\star}\big)\nonumber\\
\bm{\rho}&=&\{\mathbf{v}_1,\mathbf{v}_2\}
\end{eqnarray}

The LCM vector is $\bm{\alpha}^{\rm LCM}=(q-2,q-1)$ and the number of subpackets is equal to
\begin{eqnarray}
F(q)&=&\bm{\alpha}^{\rm LCM}\mathbf{F}^{\rm T}\nonumber\\
&=&(q-2,q-1)\left(\frac{K(q-1)}{2},\frac{K(K-q)}{2}\right)^{\rm T}\nonumber\\
&=&\frac{K(K-2)}{2}(q-1)
\end{eqnarray}
from which we see that $F(q)$ is minimized by $q^{\ast}=2$ and $F(q^{\ast})=\frac{K(K-2)}{2}$.\hfill $\square$

To illustrate the prefetching and delivery scheme above, we consider a few examples.
\begin{example}
\label{example_q2_equal_grp}
(\emph{Equal Grouping Design with Even $\bar{t}$})
For $(K,\bar{t})=(2m,2)$ with $m\geq \bar{t}=2$, consider the node grouping $\mathbf{q}=(2,2,\cdots,2)$. In this case, $(S,V)=(1,2)$. The multicasting group types, packet types and involved packet type sets are
\begin{eqnarray}
\mathbf{v}_1&=&\big(\mathbf{2}^{(m-1)},\mathbf{0}^{(1)}\big)\nonumber\\
\mathbf{v}_2&=&\big(\mathbf{2}^{(m-2)},\mathbf{1}^{(2)}\big)\nonumber\\
\mathbf{s}&=&\big(\mathbf{2}^{(m-1)},\mathbf{1}^{(1)\star}\big),\;
\bm{\rho}=\{\mathbf{v}_1,\mathbf{v}_2\}\nonumber
\end{eqnarray}

Since there is only one multicasting type, the overall LCM vector equals the local further splitting ratio vector, {\em i.e.}, $\bm{\alpha}^{\rm LCM}=(0,1)$ ($\mathbf{v}_1$ is excluded). Hence, the number of subpakcets is equal to
\begin{eqnarray}
F=\bm{\alpha}^{\rm LCM}\mathbf{F}^{\rm T}&=&(0,1)[F(\mathbf{v}_1),F(\mathbf{v}_2)]^{\rm T}\nonumber\\
&=&\frac{K(K-2)}{2}
\end{eqnarray}
and $\frac{F}{F^{\rm JCM}}=\frac{1}{K-1}$, implying an order reduction on the subpacketization level compared to the JCM scheme.

\mj{[This is the case that $t = K-2$, which was shown in [13], right?]}\xz{Yes. It is identical to the PDA design.} Next we consider the detailed caching and delivery procedure for a system with $K=8,N=4$ and $M=3$ with $t=6$. Denote $Z_k$ ($k\in[8]$) as the set of subpackets cached by node $k$. Consider a specific node-group assignment under node grouping $\mathbf{q}=(2,2,2,2)$, {\em i.e.}, $\mathcal{Q}_1=\{1,2\},\mathcal{Q}_2=\{3,4\},\mathcal{Q}_3=\{5,6\}$ and $\mathcal{Q}_4=\{7,8\}$. The set of type-$\mathbf{v}_1$ and type-$\mathbf{v}_2$ packets are

1) {Type-$\mathbf{v}_1$ Packets}
\begin{eqnarray}
\{W_{n,[K]\backslash \mathcal{Q}_i}:i\in[4],n\in[4]\}\nonumber
\end{eqnarray}

2) {Type-$\mathbf{v}_2$ Packets}
\begin{eqnarray}
\left\{W_{n,[K]\backslash \{k_1,k_2\}} :\left\lceil\frac{k_1}{2}\right\rceil \neq \left\lceil\frac{k_2}{2}\right\rceil,n\in[4]\right\}\nonumber
\end{eqnarray}

Since we have $\bm{\alpha}^{\rm LCM}=(0,1)$, which means that type-$\mathbf{v}_1$ packets are excluded and type-$\mathbf{v}_2$ packets do not need to be further split. Hence, $F=F(\mathbf{v}_2)=\binom{4}{2}\binom{2}{1}^2=24$ while $F^{\rm JCM}=168$. The cached content of node $k$ is given by
\begin{eqnarray}
Z_k&=&Z_k^{(1)}\cup Z_k^{(2)}\nonumber
\end{eqnarray}
where
\begin{eqnarray}
Z_k^{(1)}&=&\left\{W_{n,[K]\backslash\{k_1,k_2\}}: k_1,k_2\notin \mathcal{Q}_{\lceil\frac{k}{2}\rceil},\left\lceil \frac{k_1}{2}\right\rceil \neq \left\lceil \frac{k_2}{2}\right\rceil  \right\}\nonumber\\
Z_k^{(2)}&=&\left\{W_{n,[K]\backslash \left(\mathcal{Q}_{\lceil \frac{k}{2}\rceil}\cup \{ k_1 \} \right) }:k_1\notin \mathcal{Q}_{\lceil\frac{k}{2}\rceil}  \right\}\nonumber
\end{eqnarray}
from which we see that each node caches $|Z_k|=MF=72$ packets in total, satisfying the memory constraint.

Consider the multicasting group $\mathcal{S}=[K]\backslash\{8\}$ in which node 7 is selected as the transmitter and transmits the coded multicast message $\bigoplus_{k\in[6]}W_{d_k,[K]\backslash\{k,8\}}$ to the other six nodes. With the help of their cached contents, each node $k$ is able to recover the desired packet $W_{d_k,[K]\backslash\{k,8\}}$. The delivery procedure is similar for other multicasting groups. Since each coded message is useful for $t=6$ nodes, the rate is the same as the JCM scheme. \hfill $\lozenge$
\end{example}
\begin{example}
\label{example_q3_equal_grp_odd_tbar}
(\emph{Equal Grouping Design with Odd $\bar{t}$})
For $(K,\bar{t})=(3m,3)$ with $m\geq \bar{t}=3$, consider the node grouping $\mathbf{q}=(3,3,\cdots,3)$. In this case, $(S,V)=(2,3)$. The multicasting group types, packet types and involved packet type sets are
\begin{eqnarray}
\mathbf{v}_1&=&\big(\mathbf{3}^{(m-3)},\mathbf{2}^{(3)}\big)\nonumber\\
\mathbf{v}_2&=&\big(\mathbf{3}^{(m-2)},\mathbf{2}^{(1)},\mathbf{1}^{(1)}\big)\nonumber\\
\mathbf{v}_3&=&\big(\mathbf{3}^{(m-1)},\mathbf{0}^{(1)}\big)\nonumber\\
\mathbf{s}_1&=&\big(\mathbf{3}^{(m-1)},\mathbf{1}^{(1)\star}\big),\;\bm{\rho}_1=\{\mathbf{v}_2,\mathbf{v}_3\}\nonumber\\
\mathbf{s}_2&=&\big(\mathbf{3}^{(m-2)},\mathbf{2}^{(2)\star}\big),\;\bm{\rho}_2=\{\mathbf{v}_1,\mathbf{v}_2\}\nonumber\nonumber
\end{eqnarray}
and the further splitting ratio table is
\begin{center}
\begin{tabular}{|c| c c c|}
\hline
 &  $\mathbf{v}_1$ & $\mathbf{v}_2$ & $\mathbf{v}_3$  \\
\hline\hline
$\mathbf{s}_1$ &   &1 &0  \\
\hline
$\mathbf{s}_2$ &4 & 3 &\\
 \hline
 $\bm{\alpha}^{\rm LCM}$ &4 & 3 &0\\
 \hline
\end{tabular}
\end{center}
from which we obtain $\bm{\alpha}^{\rm LCM}=(4,3,0)$, implying that $\mathbf{v}_3$ is excluded (\emph{raw packet saving gain}) and $\mathbf{v}_1,\mathbf{v}_2$ have further splitting ratios of $4,3$ respectively (\emph{further splitting ratio gain}). Hence, the number of subpackets is equal to
\begin{eqnarray}
F&=&(4,3,0)[F(\mathbf{v}_1),F(\mathbf{v}_2),F(\mathbf{v}_3)]^{\rm T}\nonumber\\
&=&\frac{K(K-3)(2K-3)}{3}
\end{eqnarray}

Since we require that $K\geq 3\bar{t}=9$, the following bound holds
\begin{eqnarray}
\frac{F}{F^{\rm JCM}}=\frac{4(K-1.5)}{(K-1)(K-2)}\leq 0.5357,\;\forall K\geq 9
\end{eqnarray}
and $\frac{F}{F^{\rm JCM}}\approx \frac{4}{K}$ for large enough $K$, which implies that order reduction can be achieved.

\mj{[This example was not shown anywhere, right? If yes, can you create a table like Table 1 for this example. I think we need to include this example in ISIT.]} \xz{Yes. The corresponding table is show in Example \ref{example_q3_equal_grp_odd_tbar}.}Next consider the detailed caching and delivery procedure for a system with $K=9,N=3,M=2$ and $t=6$. Under the node grouping $\mathbf{q}=(3,3,3)$, consider a specific assignment $\mathcal{Q}_1=\{1,2,3\},\mathcal{Q}_2=\{4,5,6\}$ and $\mathcal{Q}_3=\{7,8,9\}$. There are three types of raw packets

1) {Type-$\mathbf{v}_1$ Packets}
\begin{eqnarray}
\left\{W_{n,[K]\backslash \{k_1,k_2,k_3\}}:{\scriptstyle \left(\left\lceil\frac{k_1}{3}\right\rceil,\left\lceil\frac{k_2}{3}\right\rceil,\left\lceil\frac{k_3}{3}\right\rceil\right)}=(1,2,3),n\in[3]  \right \}\nonumber
\end{eqnarray}

2) {Type-$\mathbf{v}_2$ Packets}
\begin{eqnarray}
\left\{W_{n,[K]\backslash \{k_1,k_2,k_3\}}: \left\lceil\frac{k_1}{3}\right\rceil=\left\lceil\frac{k_2}{3}\right\rceil \neq \left\lceil\frac{k_3}{3}\right\rceil ,n\in[3]    \right\}\nonumber
\end{eqnarray}

3) {Type-$\mathbf{v}_3$ Packets}
\begin{eqnarray}
\left\{W_{n,[K]\backslash \mathcal{Q}_i}:i\in[3],n\in[3]   \right\}\nonumber
\end{eqnarray}

Since $\bm{\alpha}^{\rm LCM}=(4,3,0)$, type-$\mathbf{v}_3$ packets are excluded and type-$\mathbf{v}_1$ and type-$\mathbf{v}_2$ need to be further split into 4 and 3 subpackets respectively. Hence, we have $F=4F(\mathbf{v}_1)+3F(\mathbf{v}_2)=270$ while $F^{\rm JCM}=588$. The subpackets involved in the PTB design are type-$\mathbf{v}_1$ subpackets $\{ W_{n,\mathcal{T}}^{(j)}:j=1,2,3,4\}$ and type-$\mathbf{v}_2$ subpackets $\{ W_{n,\mathcal{T}}^{(j)}:j=1,2,3\}$. The cached content of node $k$ ($k\in[9]$) is composed of type-$\mathbf{v}_1$ subpackets $Z_k^{(1)}$ and type-$\mathbf{v}_2$ subpackets $Z_k^{(2)}$, {\em i.e.},
\begin{eqnarray}
Z_k^{(1)}&=&\{W_{n,\mathcal{T}}^{(j)}:k\in\mathcal{T},j=1,2,3,4\}\nonumber\\
Z_k^{(2)}&=&\{W_{n,\mathcal{T}}^{(j)}:k\in\mathcal{T},j=1,2,3\}\nonumber
\end{eqnarray}
and $Z_k=Z_k^{(1)}\cup Z_k^{(2)}$.

There are two types of multicasting types. For the type-$\mathbf{s}_1$ multicasting groups, we consider a specific multicasting group $\mathcal{S}_1=\mathcal{Q}_1\cup\mathcal{Q}_2\cup\{7\}$, in which node 7 is selected as the only transmitter. Node 7 transmits three coded multicast messages $\bigoplus_{k\in[6]}W_{d_k,\mathcal{S}_1\backslash \{k\}}^{(j)}(j=1,2,3)$ to nodes in $\mathcal{Q}_1$ and $\mathcal{Q}_2$. Each node $k$ recovers its desired subpackets $\{W_{d_k,\mathcal{S}_1\backslash \{k\}}^{(j)}:j=1,2,3\}$ with the help of the cached content. Node 7 only transmits but receives nothing. The transmission procedure for other type-$\mathbf{s}_1$ multicasting groups is similar. Note that only type-$\mathbf{v}_2$ subpackets are involved in this transmission of type-$\mathbf{s}_1$ multicasting groups. For example, node 4 receives the following set of 18 type-$\mathbf{v}_2$ subpackets $\{W_{d_4,\mathcal{S}_1\backslash\{4\}}^{(j)}:j\in[3],n\in[3]\}$ in this delivery procedure.

For the type-$\mathbf{s}_2$ multicasting groups, we consider a specific multicasting group $\mathcal{S}_2=\mathcal{Q}_1\cup\{4,5,7,8\}$, in which nodes $4,5,7,8$ are selected as transmitters. Each node $k\in\{4,5,7,8\}$ transmits a coded multicast message to all the remaining nodes $\mathcal{S}_2\backslash \{k\}$. Denote $W=\bigoplus_{k'\in\mathcal{Q}_1} W_{d_{k',\mathcal{S}_2\backslash\{k'\}}}^{(g(k))}$ where $g:\{4,5,7,8\}\mapsto \{1,2,3,4\}$ is a bijective mapping. The transmitted message of each transmitter is shown in Table \ref{table_transmitted_message}. It can be easily verified that each node $k\in\mathcal{Q}_1$ gets a set of type-$\mathbf{v}_1$ subpackets $\{W_{d_k,\mathcal{S}_2\backslash \{k\}}^{(j)}:j=1,2,3,4\}$ and each node $k\in\{4,5,7,8\}$ gets a set of type-$\mathbf{v}_2$ subpackets $\{W_{d_k,\mathcal{S}_2\backslash \{k\}}^{(j)}:j=1,2,3\}$ after the transmission. The transmission procedure is similar for other type-$\mathbf{s}_2$ multicasting groups and it can verified that all nodes' requests can be satisfied after the whole transmission procedure.

Since within the transmission of any multicasting groups, either type-$\mathbf{s}_1$ or type-$\mathbf{s}_2$, each coded multicast message is simultaneously useful for $t$ nodes, the rate of the PTB design is the same as the JCM scheme while the number of subpackets required has been significantly reduced.\hfill $\lozenge$

\begin{table*}
\caption{Coded Multicast Messages Transmitted by Nodes 4,5,7,8. }
\begin{center}
\begin{tabular}{|c| c  |}
\hline
  Transmitter index& Coded message\\
 \hline\hline
 4  & $W_{d_8,\mathcal{S}_2\backslash\{8\}}^{(1)}\oplus W_{d_5,\mathcal{S}_2\backslash\{5\}}^{(1)}\oplus W_{d_7,\mathcal{S}_2\backslash\{7\}}^{(1)}\oplus W$ \\
 \hline
5&$W_{d_8,\mathcal{S}_2\backslash\{8\}}^{(2)}\oplus W_{d_4,\mathcal{S}_2\backslash\{4\}}^{(1)}\oplus W_{d_7,\mathcal{S}_2\backslash\{7\}}^{(2)}\oplus W$  \\
 \hline
7&$W_{d_8,\mathcal{S}_2\backslash\{8\}}^{(3)}\oplus W_{d_5,\mathcal{S}_2\backslash\{5\}}^{(2)}\oplus W_{d_4,\mathcal{S}_2\backslash\{4\}}^{(2)}\oplus W$\\
 \hline
 8&$W_{d_5,\mathcal{S}_2\backslash\{5\}}^{(3)}\oplus W_{d_4,\mathcal{S}_2\backslash\{4\}}^{(3)}\oplus W_{d_7,\mathcal{S}_2\backslash\{7\}}^{(3)}\oplus W$ \\
\hline
\end{tabular}
\end{center}\label{table_transmitted_message}
\end{table*}
\end{example}

\begin{remark}
Note that in this example, there is another choice of transmitters in the second multicasting group type, {\em i.e.}, $\mathbf{s}_2=\left(3^{\star},2,2\right)$, which will result in a smaller LCM vector $\bm{\alpha}^{\rm LCM}=(2,3,0)$ and $F=(2,3,0)[F(\mathbf{v}_1),F(\mathbf{v}_2),F(\mathbf{v}_3)]^{\rm T}=216<288$. However, such a selection of transmitters will result in a further splitting ratio linear to $K$ when the number of groups $m=\frac{K}{3}$ increases and the order reduction of subpacketization will not be preserved.
\end{remark}

The following example shows that order reduction of subpacketization can also be achieved under unequal node grouping. Note that under equal grouping design, since there is only one unique group, all the nodes cache the same number of subpackets, which implies that the node memory constraints are automatically satisfied. However, under unequal grouping, there are multiple unique groups, and nodes in different unique groups may cache different number of subpackets. As a result, the further splitting ratio vector must be designed such that the memory constraints can be satisfied, {\em i.e.}, $\bm{\alpha}^{\rm JCM}\Delta\mathbf{F}_i=0,\forall i\in[1:N_{\rm d}-1]$ where $N_{\rm d}$ denotes the number of unique groups.
\begin{example}
\label{example_unequal_design}
(\emph{Unequal Grouping Design}) For $(K,\bar{t})=(2m+1,2)(m\geq 3)$, consider the node grouping $\mathbf{q}=(3,2,2,\cdots,2)$ with $m$ groups. We have $(S,V)=(2,4)$ and the multicasting gorup and packet types are
\begin{eqnarray}
\mathbf{v}_1&=&\big(\mathbf{1}^{(1)},\mathbf{2}^{(m-1)}\big)\nonumber\\
\mathbf{v}_2&=&\big(\mathbf{2}^{(1)},\mathbf{2}^{(m-2)},\mathbf{1}^{(1)}\big)\nonumber\\
\mathbf{v}_3&=&\big(\mathbf{3}^{(1)},\mathbf{2}^{(m-3)},\mathbf{1}^{(2)}\big)\nonumber\\
\mathbf{v}_4&=&\big(\mathbf{3}^{(1)},\mathbf{2}^{(m-2)},\mathbf{0}^{(1)}\big)\nonumber\\
\mathbf{s}_1&=&\big(\mathbf{2}^{(1)\star},\mathbf{2}^{(m-1)}\big),\;\bm{\rho}_1=\{\mathbf{v}_1,\mathbf{v}_2\}\nonumber\\
\mathbf{s}_2&=&\big(\mathbf{3}^{(1)},\mathbf{2}^{(m-2)},\mathbf{1}^{(1)\star}\big),\;\bm{\rho}_2=\{\mathbf{v}_2,\mathbf{v}_3,\mathbf{v}_4\}\nonumber
\end{eqnarray}
and the further splitting ratio table is
\begin{center}
\begin{tabular}{|c| c c c c|}
\hline
 &  $\mathbf{v}_1$ & $\mathbf{v}_2$ & $\mathbf{v}_3$ &$\mathbf{v}_4$  \\
\hline\hline
$\mathbf{s}_1$ & 1  &2 & &  \\
\hline
$\mathbf{s}_2$ & & 1 &1 &0\\
 \hline
 $\bm{\alpha}^{\rm LCM}$ &1 & 2 &2 &0\\
 \hline
\end{tabular}
\end{center}
form which we obtain $\bm{\alpha}^{\rm LCM}=(1,2,2,0)$, implying that $\mathbf{v}_4$ is excluded. For other packet types, the further splitting ratio is a small constant irrespective of $K$ or $t=K-\bar{t}$. The number of subpackets is equal to
\begin{eqnarray}
F&=&(1,2,2,0)[F(\mathbf{v}_1),F(\mathbf{v}_2),F(\mathbf{v}_3),F(\mathbf{v}_4)]^{\rm T}\nonumber\\
&=&K(K-2)
\end{eqnarray}
and we have $\frac{F}{F^{\rm JCM}}=\frac{2}{K-1}$ which also implies an order gain.

Next we show that the memory constraints of nodes are satisfied under this design. There are $N_{\rm d}=2$ unique groups in $\mathbf{q}=(3,2,2,\cdots,2)$ and the number of raw packets of each type cached by nodes in the two unique groups are
\begin{eqnarray}
\mathbf{F}_1&=&[1,4(m-1),2(m-1)(m-2),m-1]\\
\mathbf{F}_2&=&[3,3(2m-3),2(m-2)^2,m-2]\\
\Delta \mathbf{F}&=&\mathbf{F}_2-\mathbf{F}_1\nonumber\\
&=&[2,2m-5,-2(m-2),-1]
\end{eqnarray}
and it can be verified that $\bm{\alpha}^{\rm LCM}\Delta \mathbf{F}^{
\rm T}=0$, implying the memory constraint is actually satisfied. \hfill $\lozenge$
\end{example}
\begin{remark}
It has been proved under the DPDA framework that $F^{\star}=K(K-2)$ is optimal for $(K,\bar{t})=(2m+1,2)$. However, the PTB design to achieve $F^{\star}$ is not unique. For example, $F^{\star}$ can be achieved by an equal-grouping PTB design. For $K=3m$ that is odd, consider the node grouping $\mathbf{q}=(3,3,\cdots,3)$ with $(S,V)=(1,2)$. The packet types and multcasting group types are $\mathbf{v}_1=\big(\mathbf{3}^{(m-2)},\mathbf{2}^{(2)}\big),\mathbf{v}_2=\big(\mathbf{3}^{(m-)},\mathbf{1}^{(1)}\big)$ and $\mathbf{s}=\big(\mathbf{3}^{(m-1)},\mathbf{2}^{(1)\star}\big),\bm{\rho}=\{\mathbf{v}_1,\mathbf{v}_2\}$. It is easy to see that $\bm{\alpha}^{\rm LCM}=(2,1)$. Hence, $F=(2,1)[F(\mathbf{v}_1),F(\mathbf{v}_2)]^{\rm T}=K(K-2)$. This new design is indeed different from the unequal grouping design since the number of required raw packets are different, indicating some structural difference. More specifically, the number of raw packets required here is equal to $F(\mathbf{v}_1)+F(\mathbf{v}_2)=\frac{K(K-1)}{2}=\binom{K}{t}$ which is due to the fact that no exclusion of raw packets occurs in this case. On the other hand, the number of raw packets required in Example \ref{example_unequal_design} is equal to $\frac{(K-1)^2}{2}+1< \binom{K}{t},\forall K\geq 9$, which is due to the exclusion of type-$\mathbf{v}_4$ raw packets. The two different designs implies some sort of \emph{trade-off between raw packet saving gain and further splitting ratio gain} under optimal subpacketization.
\end{remark}

\begin{remark}
The node grouping $\mathbf{q}=(3,2,2,\cdots,2)$ actually achieves the minimum $F$ among a set of unequal grouping designs, which is shown in the following lemma.
\end{remark}
\begin{lemma}
\label{lemma_unequal_series_min}
\emph{For $(K,\bar{t})=(2(m+q)+1,2)$ with $q\geq 1,m\geq 2$, the node grouping $\mathbf{q}=(3,2,\cdots,2)$ yields the minimum $F$ among the set of equal groupings $\{\mathbf{q}=(p,2,2,\cdots,2)(p=2q+1):q+m=\frac{K-1}{2} \}$}.
\end{lemma}

\emph{Proof}: $(S,V)=(2,4)$ and the packet types and multicasting types are (transmitters are marked by the superscript $^\star$)
\begin{eqnarray}
\mathbf{v}_1&=&\big(p-2,\mathbf{2}^{(m)}\big)\nonumber\\
\mathbf{v}_2&=&\big(p,\mathbf{2}^{(m-1)},0\big)\nonumber\\
\mathbf{v}_3&=&\big(p-1,\mathbf{2}^{(m-1)},1\big)\nonumber\\
\mathbf{v}_4&=&\big(p,\mathbf{2}^{(m-2)},\mathbf{1}^{(2)}\big)\nonumber\\
\mathbf{s}_1&=&\big((p-1)^{\star},\mathbf{2}^{(m-1)}\big),\;\bm{\rho}_1=\{\mathbf{v}_1,\mathbf{v}_3\}\nonumber\\
\mathbf{s}_1&=&\big(p,\mathbf{2}^{(m-1)},1^{\star}\big),\;\bm{\rho}_2=\{\mathbf{v}_2,\mathbf{v}_3,\mathbf{v}_4\}\nonumber
\end{eqnarray}
form which we obtain $\bm{\alpha}^{\rm LCM}=(p-2,0,p-1,p-1)$ where $\mathbf{v}_2$ is excluded. We also have
\begin{eqnarray}
\mathbf{F}&=&\left(\frac{p(p-1)}{2},m,2pm,2m(m-1)    \right)\\
\Delta\mathbf{F}&=&\left((p-1)^2,-1,2m-p,-2(m-1)\right)
\end{eqnarray}

It can be easily verified that $\bm{\alpha}^{\rm LCM}\Delta\mathbf{F}^{\rm T}=0$, which implies that the node memory constraint is satisfied. The number of subpackets is equal to
\begin{eqnarray}
F(q)=\bm{\alpha}^{\rm LCM}\mathbf{F}^{\rm T}=qK(K-2)
\end{eqnarray}
from which we can see that $F(q)$ is minimized by $q^{\ast}=1$ and $F(q^{\ast})=K(K-2)$, which completes the proof of Lemma \ref{lemma_unequal_series_min}. \hfill $\square$

\subsection{2-Group Equal Grouping Design}
\label{subsection_2_groups}
We present the achievable scheme of Theorem \ref{theorem_2_groups} in this section. The two-group node grouping PTB design achieves a (constant) subpacketization reduction more than one-half compared to the JCM scheme.

For $K=2q\geq 2(t+1)$ and $t=2r(r\in\mathbb{Z}^+)$, consider the equal grouping $\mathbf{q}=(q,q)$ with $q=\frac{K}{2}$. We have $(S,V)=(r+1,r+1)$ and
\begin{eqnarray}
\mathbf{v}_i&=&(t-i+1,i-1),\;\forall i\in[r+1]\nonumber\\
\mathbf{s}_j&=&(t-j+2,j-1^{\star}),\;\forall j\in[r+1]\nonumber\\
\bm{\rho}_1&=&\{\mathbf{v}_1\}\nonumber\\
\bm{\rho}_j&=&\{\mathbf{v}_{j-1},\mathbf{v}_j\},\;\forall j\in[2:r+1]\nonumber
\end{eqnarray}

The local further splitting ratio vectors are $\bm{\alpha}_1=(t)$ and $\bm{\alpha}_j=(j-2,j-1),\forall j\in[2:r+1]$. The further splitting ratio table is described in Table \ref{table_2_grps_general}. It is easy to derive the LCM vector since all the involved packet type sets are disjoint. We have
\begin{eqnarray}
\bm{\alpha}^{\rm LCM}=(0,1,2,\cdots,r-1,r)=[0:\frac{t}{2}]
\end{eqnarray}

\begin{table*}
\caption{Further Splitting Ratio Table for Two-Group PTB Design}
\begin{center}
\begin{tabular}{|c| c c c c c c c c c c c |}
\hline
  & $\mathbf{v}_1$ & $\mathbf{v}_2$ & $\mathbf{v}_3$ &$\cdots $ &$\mathbf{v}_{j-1}$ &$\mathbf{v}_j$ &$\mathbf{v}_{j+1}$ &$\cdots $ &$\mathbf{v}_{r-1}$&$\mathbf{v}_{r}$ & $\mathbf{v}_{r+1}$\\
 \hline\hline
  $\mathbf{s}_1$ &$t$&   &   &  &  & &   & & & & \\
 \hline
   $\mathbf{s}_2$&0 & 1 &   &  &  &  & &  & &&  \\
 \hline
   $\mathbf{s}_3$& & 1 & 2  &  &  &  &  &  && & \\
 \hline
 $\vdots$ & &  &   & $\ddots$  &  &    &&  & && \\
 \hline
  $\mathbf{s}_j$ & &  &   & & $j-2$ & $j-1$&  & &  &&  \\
 \hline
   $\mathbf{s}_{j+1}$ & &  &   &  &  &$j-1$  & $j$ && & &  \\
 \hline
   $\vdots $ & &  &   &  &  &  &  & $\ddots$ &&&   \\
 \hline
   $\mathbf{s}_r$ & &  &   &  &  &  &  &&$r-2$&  $r-1$&  \\
 \hline
  $\mathbf{s}_{r+1}$ & &  &   &  &  &  &  &  &&$r-1$ &$r$\\
  \hline
    $\bm{\alpha}^{\rm LCM}$ &0 & 1 & 2  & $\cdots$ & $j-2$ &$j-1$ &$j$  &$\cdots$ & & $r-1$ &$r$  \\
    \hline
\end{tabular}
\end{center}\label{table_2_grps_general}
\end{table*}

The number of subpackets is equal to
\begin{eqnarray}
F=\bm{\alpha}^{\rm LCM}\mathbf{F}^{\rm T}&=&\sum_{j=1}^{r+1}\alpha_j F(\mathbf{v}_j)
\end{eqnarray}
in which
\begin{eqnarray}
F(\mathbf{v}_j)&=&2\binom{q}{j-1}\binom{q}{t-j+1}\nonumber\\
&=&\frac{2\prod_{i=0}^{j-2}(q-i)\prod_{i=0}^{t-j}(q-i)}{(j-1)!(t-j+1)!}\nonumber\\
&\overset{{\rm large }\, q}{\approx}& \frac{q^t}{(j-1)!(t-j+1)!},\;\forall j\in[r]\\
F(\mathbf{v}_{r+1})&=&\binom{q}{r}^2=\frac{\prod_{i=0}^{r-1}(q-i)^2}{(r!)^2}\nonumber\\
&\overset{{\rm large }\, q}{\approx}& \frac{q^t}{\left((\frac{t}{2})!\right)^2}
\end{eqnarray}

Hence, we have the following upper bound
\begin{eqnarray}
\frac{F}{F^{\rm JCM}}&=&\frac{\sum_{j=1}^{r+1}(j-1)F(\mathbf{v}_j)}{t\sum_{j=1}^{r+1}F(\mathbf{v}_j)}\nonumber\\
&<&\frac{r\sum_{j=1}^{r+1}F(\mathbf{v}_j)}{t\sum_{j=1}^{r+1}F(\mathbf{v}_j)}=\frac{r}{t}=\frac{1}{2}
\end{eqnarray}
and assuming $r\geq 2$, {\em i.e.}, $t\ge 4$,
\begin{eqnarray}
\zeta(t)&=&\lim\limits_{q\to\infty}\frac{\sum_{j=1}^{r+1}(j-1)F(\mathbf{v}_j)}{t\sum_{i=j}^{r+1}F(\mathbf{v}_j)}\nonumber\\
&=&\lim\limits_{q\to\infty}\frac{1}{t}\frac{\sum_{j=1}^{r}(j-1)F(\mathbf{v}_j)+rF(\mathbf{v}_{r+1})}{\sum_{j=1}^{r}F(\mathbf{v}_j)+F(\mathbf{v}_{r+1})}\nonumber\\
&=&\lim\limits_{q\to\infty}\frac{1}{t}\frac{\sum_{j=2}^{r}\frac{2\prod_{i=0}^{j-2}(q-i)\prod_{i=0}^{t-j}(q-i)}{(j-2)!(t-j+1)!}+r\frac{\prod_{i=0}^{r-1}(q-i)^2}{(r!)^2}}{\sum_{j=1}^{r}\frac{2\prod_{i=0}^{j-2}(q-i)\prod_{i=0}^{t-j}(q-i)}{(j-1)!(t-j+1)!}+\frac{\prod_{i=0}^{r-1}(q-i)^2}{(r!)^2}}\nonumber\\
&=&\lim\limits_{q\to\infty}\frac{1}{t}\frac{\sum_{j=2}^{r}\frac{2q^t}{(j-2)!(t-j+1)!}+q^tr\left(r!\right)^{-2}}{\sum_{j=1}^{r}\frac{2q^t}{(j-1)!(t-j+1)!}+q^t\left(r!\right)^{-2}}\nonumber\\
&=&\frac{1}{t}\frac{\sum_{j=2}^{r}\frac{2}{(j-2)!(t-j+1)!}+r\left(r!\right)^{-2}}{\sum_{j=1}^{r}\frac{2}{(j-1)!(t-j+1)!}+\left(r!\right)^{-2}}\nonumber\\
&=&\frac{1}{t}\frac{\sum_{j=2}^{r}\frac{2(j-1)}{(j-1)!(t-j+1)!}+r\left(r!\right)^{-2}}{\sum_{j=1}^{r}\frac{2}{(j-1)!(t-j+1)!}+\left(r!\right)^{-2}}\nonumber\\
&\overset{(a)}{<}&\frac{r}{t}\frac{\sum_{j=2}^{r}\frac{2}{(j-1)!(t-j+1)!}+\left(r!\right)^{-2}}{\sum_{j=1}^{r}\frac{2}{(j-1)!(t-j+1)!}+\left(r!\right)^{-2}}\nonumber\\
&\overset{(b)}{=}&\frac{r}{t}\frac{\sigma_r}{\sigma_r+\frac{2}{t!}}\nonumber\\
&=&\frac{r}{t}\left(1+\frac{2}{t!\sigma_r}\right)^{-1}\nonumber\\
&\overset{(c)}{=}&\frac{r}{t}\left(1+\frac{2}{2^t-2}\right)^{-1}\nonumber\\
&=&\frac{1}{2}\left(1-\frac{1}{2^{t-1}}\right)
\end{eqnarray}
where $(a)$ is due to $j-1<r,\forall j\in[2:r]$. In (b), we used the notation $\sigma_r:=\sum_{j=2}^{r}\frac{2}{(j-1)!(t-j+1)!}+\left(r!\right)^{-2}$. Step $(c)$ is due to
\begin{eqnarray}
t!\sigma_r&=&t!\left(\sum_{j=2}^{r}\frac{2}{(j-1)!(t-j+1)!}+\left(r!\right)^{-2}\right)\nonumber\\
&=&2\sum_{j=2}^{r}\frac{t!}{(j-1)!(t-j+1)!}+\frac{t!}{r!(t-r)!}\nonumber\\
&=&2\sum_{i=1}^{r-1}\frac{t!}{i!(t-i)!}+\frac{t!}{r!(t-r)!}\nonumber\\
&=&2\sum_{i=1}^{r-1}\binom{t}{i}+\binom{t}{r}\nonumber\\
&=&2^t-2
\end{eqnarray}
where in the last step we used the fact that $\sum_{i=0}^{t}\binom{t}{i}=2^t$.

If $r=1$, {\em i.e.}, $t=2$, through example \ref{example_2_group_t=2}, we see that $\zeta(t)=\lim\limits_{K\to\infty}\frac{1}{4}\frac{K}{K-1}=0.25$. Hence, the proof of Theorem \ref{theorem_2_groups} is complete. \hfill $\square$

\begin{example}
\label{example_2_group_t=2}
Consider $t=2$ with $\mathbf{q}=(\frac{K}{2},\frac{K}{2})$. We have $(S,V)=(2,2)$ and
\begin{eqnarray}
\mathbf{v}_1&=&(2,0),\;\mathbf{v}_2=(1,1)\nonumber\\
\mathbf{s}_1&=&(3,0),\;\bm{\rho}_1=\{\mathbf{v}_1\}\nonumber\\
\mathbf{s}_1&=&(2,1^{\star}),\;\bm{\rho}_2=\{\mathbf{v}_1,\mathbf{v}_2\}\nonumber
\end{eqnarray}
and $\bm{\alpha}^{\rm JCM}=(0,1)$. The number of subpackets is equal to
\begin{eqnarray}
F=(0,1)[F(\mathbf{v}_1),F(\mathbf{v}_2)]^{\rm T}=\frac{K^2}{4}
\end{eqnarray}
and $\frac{F}{F^{\rm JCM}}=\frac{1}{4}\frac{K}{K-1}\leq 0.3,\forall K\geq 6$. \hfill $\lozenge$
\end{example}

\begin{example}
\label{example_2_grp_even_t=4}
Consider $t=4$ with $\mathbf{q}=(\frac{K}{2},\frac{K}{2})$. We have $(S,V)=(3,3)$ and
\begin{eqnarray}
\mathbf{v}_1&=&(4,0)\nonumber\\
\mathbf{v}_2&=&(3,1)\nonumber\\
\mathbf{v}_3&=&(2,2)\nonumber\\
\mathbf{s}_1&=&(5,0),\;\;\,\bm{\rho}_1=\{\mathbf{v}_1\}\nonumber\\
\mathbf{s}_2&=&(4,1^{\star}),\: \bm{\rho}_2=\{\mathbf{v}_2,\mathbf{v}_1\}\nonumber\\
\mathbf{s}_3&=&(3,2^{\star}),\:
\bm{\rho}_3=\{\mathbf{v}_3,\mathbf{v}_2\}\nonumber
\end{eqnarray}
and $\bm{\alpha}^{\rm LCM}=(0,1,2)$. The number of subpackets is equal to
\begin{eqnarray}
F&=&(0,1,2)[F(\mathbf{v}_1),F(\mathbf{v}_2),F(\mathbf{v}_3)]^{\rm T}\nonumber\\
&=&\frac{K^2(K-2)(5K-14)}{96}
\end{eqnarray}

Since we require that $K\geq 2(t+1)=10$, the following bound holds
\begin{eqnarray}
\frac{F}{F^{\rm JCM}}=\frac{0.3125K(K-2.8)}{(K-1)(K-3)}\leq 0.3571,\;\forall K\geq 10
\end{eqnarray}
and $\lim\limits_{K\to\infty}\frac{F}{F^{\rm JCM}}=0.3125$, which is less than one-third. \hfill $\lozenge$
\end{example}

\subsection{Heterogeneous Subpacket Size Design}
\label{subsection_heter_pkt_size}
In this section we present the achievable scheme for Theorem \ref{theorem_heter_pkt_size} via the following example.

\begin{example}
\label{example_heter_pkt_size}
Consider $(K,t)=(2q+1,2)(q\geq 3)$ with the node grouping $\mathbf{q}=(q+1,q)$. We have $(S,V)=(4,3)$. For each packet type $\mathbf{v}_j (j\in[3])$, we divide it into two sub-types $\mathbf{v}_j^{(1)}$ and $\mathbf{v}_j^{(2)}$. Subpackets of the two sub-types are cached in the same set of nodes, but have different subpacket sizes $\ell_1$ and $\ell_2$. We define a \emph{coupled group}, denoted by $\mathcal{G}$, as a set of sub-types which have the same size. In this case we consider the two coupled groups $\mathcal{G}_1=\{\mathbf{v}_2^{(1)},\mathbf{v}_3^{(1)}\}$ and $\mathcal{G}_2=\{\mathbf{v}_2^{(2)}\}$. We assume that subpackets belonging to $\mathcal{G}_1$ have size $\ell_1$ bits and subpackets belonging to $\mathcal{G}_2$ have size $\ell_2$ bits.

The three packet types are $\mathbf{v}_1=(0,2),\mathbf{v}_2=(1,1)$ and $\mathbf{v}_3=(2,0)$. We consider two different choices of transmitters (marked by the superscript $^\star$) within each multicasting type, {\em i.e.},

Choice 1:\begin{eqnarray}
\mathbf{s}_1&=&(0,3^{\star}),\mathbf{s}_2=(1^{\star},2)\nonumber\\
\mathbf{s}_3&=&(2^{\star},1),\mathbf{s}_4=(3^{\star},0)\nonumber
\end{eqnarray}

Choice 2:
\begin{eqnarray}
\mathbf{s}_1&=&(0,3^{\star}),\mathbf{s}_2=(1^{\star},2)\nonumber\\
\mathbf{s}_3&=&(2,1^{\star}),\mathbf{s}_4=(3^{\star},0)\nonumber
\end{eqnarray}

The above two choices lead to two different further splitting ratio tables. The further splitting ratio table $\mathbf{\Lambda}_1$ corresponding to the first choice is
\begin{center}
\begin{tabular}{|c| c c c |}
\hline
 &  $\mathbf{v}_1$ & $\mathbf{v}_2$ & $\mathbf{v}_3$  \\
\hline\hline
$\mathbf{s}_1$ & 2  & &   \\
\hline
$\mathbf{s}_2$ & 0& 1 & \\
 \hline
 $\mathbf{s}_3$ & & 1 &2 \\
 \hline
 $\mathbf{s}_4$ & &  &2 \\
 \hline
 $\bm{\alpha}^{(1)}$ &0 & 1 &2 \\
 \hline
\end{tabular}
\end{center}
and the LCM vector corresponding to $\mathcal{G}_1$ is $\bm{\alpha}^{(1)}=(0,1,2)$, which implies that sub-type $\mathbf{v}_1^{(1)}$ is excluded and $\mathbf{v}_2^{(1)}$ and $\mathbf{v}_3^{(1)}$ are split into one and two subpackets respectively.

The further splitting ratio table $\mathbf{\Lambda}_2$ corresponding to the second choice is
\begin{center}
\begin{tabular}{|c| c c c |}
\hline
 &  $\mathbf{v}_1$ & $\mathbf{v}_2$ & $\mathbf{v}_3$  \\
\hline\hline
$\mathbf{s}_1$ & 2  & &   \\
\hline
$\mathbf{s}_2$ & 0& 1 & \\
 \hline
 $\mathbf{s}_3$ & & 1 &0 \\
 \hline
 $\mathbf{s}_4$ & &  &2 \\
 \hline
 $\bm{\alpha}^{(2)}$ &0 & 1 &0 \\
 \hline
\end{tabular}
\end{center}
and the LCM vector corresponding to $\mathcal{G}_2$ is $\bm{\alpha}^{(2)}=(0,1,0)$, which implies that sub-type $\mathbf{v}_1^{(2)},\mathbf{v}_3^{(2)}$ are both excluded and $\mathbf{v}_2^{(2)}$ does not need to be further split. As a result, the overall LCM vector is equal to
\begin{eqnarray}
\bm{\alpha}^{\rm LCM}=\bm{\alpha }^{(1)}+\bm{\alpha }^{(2)}=(0,2,2)
\end{eqnarray}

The following notations are used. For $i,j\in[2]$, $\mathbf{F}_j^{(i)}=[F_j(\mathbf{v}_1^{(i)}),F_j(\mathbf{v}_2^{(i)}),F_j(\mathbf{v}_3^{(i)})],\mathbf{F}^{(i)}=[F(\mathbf{v}_1^{(i)}),F(\mathbf{v}_2^{(i)}),F(\mathbf{v}_3^{(i)})]$ and $\Delta\mathbf{F}^{(i)}=\mathbf{F}_2^{(i)}-\mathbf{F}_1^{(i)}$. Then the memory constraint can be represented as:
\begin{eqnarray}
\ell_1\Delta\mathbf{F}^{(1)}{\bm{\alpha}^{(1)}}^{\rm T}+\ell_2\Delta\mathbf{F}^{(2)}{\bm{\alpha}^{(2)}}^{\rm T}=0
\end{eqnarray}
which implies that
\begin{eqnarray}
\gamma:=\frac{\ell_1}{\ell_2}&=&-\frac{\Delta\mathbf{F}^{(2)}{\bm{\alpha}^{(2)}}^{\rm T}}{\Delta\mathbf{F}^{(1)}{\bm{\alpha}^{(1)}}^{\rm T}}\nonumber\\
&=&-\frac{\left(\mathbf{F}_2^{(2)}-\mathbf{F}_1^{(2)}\right){\bm{\alpha}^{(2)}}^{\rm T}}{\left(\mathbf{F}_2^{(1)}-\mathbf{F}_1^{(1)}\right){\bm{\alpha}^{(1)}}^{\rm T}}\nonumber\\
&=&-\frac{\left[\binom{q-1}{1},\binom{q+1}{1}-\binom{q}{1},-\binom{q}{1}\right](0,1,0)^{\rm T}}{\left[\binom{q-1}{1},\binom{q+1}{1}-\binom{q}{1},-\binom{q}{1}\right](0,1,2)^{\rm T}}\nonumber\\
&=&-\frac{\binom{q+1}{1}-\binom{q}{1}}{\binom{q+1}{1}-3\binom{q}{1}}\nonumber\\
&=&\frac{1}{2q-1}\nonumber\\
&=&\frac{1}{K-2}\in\mathbb{Q}^+
\end{eqnarray}

Denote $F_1=\mathbf{F}^{(1)}{\bm{\alpha}^{(1)}}^{\rm T}$ and $F_2=\mathbf{F}^{(2)}{\bm{\alpha}^{(2)}}^{\rm T}$. The node cache memory constraint also requires
\begin{eqnarray}
\ell_1\mathbf{F}_1^{(1)}{\bm{\alpha}^{(1)}}^{\rm T}+\ell_2\mathbf{F}_1^{(2)}{\bm{\alpha}^{(2)}}^{\rm T}=M(\ell_1F_1+\ell_2F_2)\quad {\rm bits}
\end{eqnarray}
which is equivalent to
\begin{eqnarray}
M(\ell_1F_1+\ell_2F_2)&=&\gamma\ell_2\mathbf{F}_1^{(1)}{\bm{\alpha}^{(1)}}^{\rm T}+\ell_2\mathbf{F}_1^{(2)}{\bm{\alpha}^{(2)}}^{\rm T}\nonumber\\
&=&\ell_2\left(\gamma\mathbf{F}_1^{(1)}{\bm{\alpha}^{(1)}}^{\rm T}+\mathbf{F}_1^{(2)}{\bm{\alpha}^{(2)}}^{\rm T}\right)\nonumber
\end{eqnarray}

Hence, we have
\begin{eqnarray}
\ell_1=\gamma\ell_2&=&\frac{\gamma M(\ell_1F_1+\ell_2F_2)}{\gamma\mathbf{F}_1^{(1)}{\bm{\alpha}^{(1)}}^{\rm T}+\mathbf{F}_1^{(2)}{\bm{\alpha}^{(2)}}^{\rm T}}\\
&=&\frac{\gamma M(\ell_1F_1+\ell_2F_2)}{\gamma\mathbf{F}_2^{(1)}{\bm{\alpha}^{(1)}}^{\rm T}+\mathbf{F}_2^{(2)}{\bm{\alpha}^{(2)}}^{\rm T}}
\end{eqnarray}
which specifies the number of bits contained in the subpackets in both coupled groups $\mathcal{G}_1$ and $\mathcal{G}_2$.

The number of subpackets require for this design is equal to
\begin{eqnarray}
F&=&F_1+F_2\nonumber\\
&=&\mathbf{F}^{(1)}{\bm{\alpha}^{(1)}}^{\rm T}+\mathbf{F}^{(2)}{\bm{\alpha}^{(2)}}^{\rm T}\nonumber\\
&\overset{(a)}{=}&\mathbf{F}^{(1)}(\bm{\alpha}^{(1)}+\bm{\alpha}^{(2)})^{\rm T}\nonumber\\
&=&\mathbf{F}^{(1)}{\bm{\alpha}^{\rm LCM}}^{\rm T}\nonumber\\
&=&\left[\binom{q}{2},\binom{q+1}{1}\binom{q}{1},\binom{q+1}{2}  \right](0,2,2)^{\rm T}\nonumber\\
&=&\frac{3(K^2-1)}{4}
\end{eqnarray}
where in step $(a)$ we used the fact that $\mathbf{F}^{(1)}=\mathbf{F}^{(2)}$. We also have $\frac{F}{F^{\rm JCM}}=\frac{3}{4}(1+\frac{1}{K})\leq \frac{6}{7},\forall K\geq 7$. The asymptotic constant subpacketization reduction is $\lim_{K\to\infty}\frac{F}{F^{\rm JCM}}=\frac{3}{4}$, which meets the upper bound on $\zeta(t)$ in Theorem \ref{theorem_heter_pkt_size}.
\hfill  $\lozenge$
\begin{remark}
In this heterogeneous subpacket size design, the constant subpacketization reduction comes from the exclusion of the packet type $\mathbf{v}=(0,2)$ which is formed  by selecting $t=2$ nodes within the second node group containing $\frac{K-1}{2}$ nodes. Only raw packet saving gain is available in the above example since $\bm{\alpha}^{\rm LCM}=(0,2,2)$. However, for even $t\geq 4$, both raw packet saving gain and further splitting ratio gain are available due to $\bm{\alpha}^{\rm LCM}=(0,2,4,\cdots,t-2,t,t,\cdots,t)$.
\end{remark}

\begin{remark}
The transmissions for each coupled group are separated since only sub-types of subpackets belonging to the same coupled group can form multicasting groups. As a result, the overall further splitting ratio vector will be the summation of the individual splitting ratio vectors for each coupled group. The heterogenity of subpacket size among coupled groups enforces the satisfaction of node memory constraints under unequal node grouping.
\end{remark}

\begin{remark}
The node grouping $\mathbf{q}=(q+1,q)$ actually achieves the minimum $F$ over a set of nodes groupings for a fixed $K=2q+1$ under the heterogeneous subpacket design, which is shown in the following lemma.
\end{remark}

\begin{lemma}
\label{lemma_heter_series}
\emph{For $(K,t)=(2q+1,2)(q\geq 3)$, the node grouping $\mathbf{q}=(q+1,q)$ achieves the minimum $F$ among the set of node groupings $\{\mathbf{q}=(q_1,K-q_1):\frac{K+1}{2}\leq q_1\leq K-3 \}$}.
\end{lemma}

\emph{Proof}: Note that the further splitting ratio table is exactly as that of Example \ref{example_heter_pkt_size}. The two coupled groups are $\mathcal{G}_1$ and $\mathcal{G}_2$ are also the same. Hence, we have $\bm{\alpha}^{(1)}=(0,1,0),\bm{\alpha}^{(2)}=(0,1,2)$ and $\bm{\alpha}^{\rm LCM}=(0,2,2)$. To satisfy the node memory constraint, the corresponding subpacket length ratio must be
\begin{eqnarray}
\gamma(q_1)=\frac{\ell_1}{\ell_2}&=&-\frac{\Delta\mathbf{F}^{(2)}{\bm{\alpha}^{(2)}}^{\rm T}}{\Delta\mathbf{F}^{(1)}{\bm{\alpha}^{(1)}}^{\rm T}}\nonumber\\
&=&-\frac{\left(K-q_1-1,2q_1-K,1-q_1\right)(0,1,0)^{\rm T}}{\left(K-q_1-1,2q_1-K,1-q_1\right)(0,1,2)^{\rm T}}\nonumber\\
&=&\frac{2q_1-K}{K-2}
\end{eqnarray}
where $\frac{1}{K-2}\leq\gamma(q_1) \leq \frac{K-6}{K-2},\forall q_1\in[\frac{K+1}{2}:K-3]$.

The number of subpacket is equal to
\begin{eqnarray}
F(q_1)&=&(0,2,2)\left[\binom{K-q_1}{2},\binom{q_1}{1}\binom{K-q_1}{1},\binom{q_1}{2}\right]^{\rm T}\nonumber\\
&=&q_1(2K-1-q_1)
\end{eqnarray}
form which we can see that $F(q_1)$ is minimized by $q_1^{\ast}=\frac{K+1}{2}=q+1$ and $F(q_1^{\ast})=\frac{3(K^2-1)}{4}$, which completes the proof of lemma \ref{lemma_heter_series}.  \hfill $\square$
\end{example}

%
%

\section{Discussion And Future Work}
\label{section_conclusion}
In this paper, we focused on the subpacketization reduction problem for D2D coded caching while achieving the optimal rate. We first developed the Packet Type-based (PTB) design framework where the subpacketization reduction problem can be formulated as an integer optimization problem with node memory constraints and proper restrictions on the set of candidate further splitting ratio vectors. Each feasible solution of the optimization problem corresponds to a valid PTB design of caching schemes. Then we focused on some special cases of node grouping, {\em i.e.}, equal grouping method and certain unequal grouping methods, and proposed several classes of PTB design with constant or order reduction of subpacketization which comes from a combination of the raw packet saving gain and the further splitting ratio gain. The result showed that the previously well-known JCM scheme is not optimal in terms of subpacketization in general. We also came up with the concept of heterogeneous subpacket size to deal with node memory satisfaction under unequal node goruping. Since the exhaustive search for the optimal solution is almost impossible for large system parameters $K$ and $t$ due to the complex nature of partitions, the properties of the integer optimization problem and the characterization of the optimal solution $F^{\ast}$ need to be further studied. More specifically, the following issues need to be considered:
\subsection{Equal Node Grouping}
In theorem \ref{theorem_1_large_t} we only characterized the order reduction of subpacketization for even values of $\bar{t}:=K-t$ under the node grouping $\mathbf{q}=(2,2,\cdots,2)$. It is still not clear whether a node grouping $\mathbf{q}=(q,q,\cdots,q)$ with $q\geq 3$ can achieve even lower subpacketization level even though we can show that for $\bar{t}=2$ the node grouping $\mathbf{q}=(2,2,\cdots,2)$ actually achieves the best subpacketization among the set of node groupings $\{\mathbf{q}=(q,q,\cdots,q):q\geq 2\}$.

\subsection{Unequal Node Grouping}
For $K=2(m+q)+1,\bar{t}=2$, we have proved the node grouping $\mathbf{q}=(q_1,2,2,\cdots,2)$ with $q_1=2q+1$ satisfies the memory constraint when all the subpackets have the same size. We also introduced the heterogeneous subpacket size design to enforce raw packet saving under node grouping $\mathbf{q}=(q+1,1)$ for $K=2q+1$. However, it is not clear whether other unequal node groupings exists such that subpacketization reduction can be achieved, especially those with three or more different unique groups. Hence, it is meaningful to find out more unequal node groupings that achieves good reduction performance for general values of $t$.

\begin{appendices}
\section{Trivial PTB Design}
A direct result of Corollary \ref{corollary_exclusion_small_t} is the trivial PTB design which is formally described as follows.

For $K=mq(q\geq t)$ with the node grouping $\mathbf{q}=(q,q,\cdots,q)$. Corollary \ref{corollary_exclusion_small_t} indicates that the packet type $\mathbf{v}=(q,q,\cdots,q-t)$ can be excluded. It is trivial to mention that the LCM vector
\begin{eqnarray}
\bm{\alpha}^{\rm LCM}=(0,\underbrace{t,t,\cdots,t}_{V-1\,{\rm terms}})
\end{eqnarray}
is always feasible. As a result, the number of subpackets is equal to
\begin{eqnarray}
F=\bm{\alpha}^{\rm LCM}\mathbf{F}^{\rm T}&=&\left(\bm{\alpha}^{\rm JCM}-(\bm{\alpha}^{\rm JCM}-\bm{\alpha}^{\rm LCM})  \right)\mathbf{F}^{\rm T}\nonumber\\
&=&\bm{\alpha}^{\rm JCM}\mathbf{F}^{\rm T}-(t,0,0,\cdots,0)\mathbf{F}^{\rm T}\nonumber\\
&=&F^{\rm JCM}-tF(\mathbf{v})\nonumber\\
&=&t\binom{K}{t}-mt\binom{q}{t}
\end{eqnarray}
and
\begin{eqnarray}
\frac{F}{F^{\rm JCM}}&=&\frac{t\binom{K}{t}-mt\binom{q}{t}}{t\binom{K}{t}}\nonumber\\
&=&1-\frac{m\binom{q}{t}}{\binom{K}{t}}\nonumber\\
&=&1-\frac{\frac{mq!}{(q-t)!t!}}{\frac{K!}{(K-t)!t!}}\nonumber\\
&=&1-\frac{mq!(K-t)!}{K!(q-t)!}\nonumber\\
&=&1-\frac{m\left(\frac{K}{m}\right)!(K-t)!}{K!(\frac{K}{m}-t)!}\nonumber\\
&=&1-\frac{m\prod_{i=0}^{t-1}\left(\frac{K}{m}-i\right)!}{\prod_{i=0}^{t-1}(K-i)}
\end{eqnarray}
and for large enough $K$ and constant $t$, we have
\begin{eqnarray}
\lim\limits_{K\to\infty}\frac{F}{F^{\rm JCM}}&=&\lim\limits_{K\to\infty}\left(1-\frac{m\prod_{i=0}^{t-1}\left(\frac{K}{m}-i\right)!}{\prod_{i=0}^{t-1}(K-i)}\right)\nonumber\\
&=&1-\frac{m\left(\frac{K}{m}\right)^{t}}{K^t}\nonumber\\
&=&1-\frac{1}{m^{t-1}}
\end{eqnarray}

\begin{remark}
In the trivial PTB design only raw packet saving gain is available since all the non-zero further splitting ratios are equal to $t$. From the asymptotic analysis we see that the raw packet saving gain $\frac{1}{m^{t-1}}$ decreases exponentially as $t$ grows for a fixed number of groups $m$. This subpacketization reduction is almost negligible for large $t$.
\end{remark}

\section{Supplemental Example of Order Reduction}
In Theorem \ref{theorem_1_large_t} we showed that order reduction of subpacketization can be achieved under the node grouping $\mathbf{q}=(2,2,\cdots,2)$ for large values of $K$ and a constant $\bar{t}$. However, there are other node groupings that also achieves order reduction for large $K$. We will present several more results on the set of possible node groupings which facilitate order reduction in the journal version. The following example shows that order reduction can be achieved under $\mathbf{q}=(\bar{t},\bar{t},\cdots,\bar{t})$ with $K=m\bar{t}$ and $\bar{t}=K-t$.

\begin{example}
For $(K,\bar{t})=(4m,4)(m\geq 4)$, consider the node grouping $\mathbf{q}=(4,4,\cdots,4)$. We have $(S,V)=(3,5)$. The corresponding packet types and multicasting types are (transmitters are marked by the superscript $^{\star}$)
\begin{eqnarray}
\mathbf{v}_1&=&\big(\mathbf{4}^{(m-4)},\mathbf{3}^{(4)}\big)\nonumber\\
\mathbf{v}_2&=&\big(\mathbf{4}^{(m-3)},2,\mathbf{1}^{(2)}\big)\nonumber\\
\mathbf{v}_3&=&\big(\mathbf{4}^{(m-2)},\mathbf{2}^{(2)}\big)\nonumber\\
\mathbf{v}_4&=&\big(\mathbf{4}^{(m-2)},3,1\big)\nonumber\\
\mathbf{v}_5&=&\big(\mathbf{4}^{(m-1)},0\big)\nonumber\\
\mathbf{s}_1&=&\big(\mathbf{4}^{(m-3)},\mathbf{3}^{(3)\star}\big),\;\bm{\rho}_1=\{\mathbf{v}_1,\mathbf{v}_2\}\nonumber\\
\mathbf{s}_2&=&\big(\mathbf{4}^{(m-2)},3,2^{\star}\big),\;\bm{\rho}_2=\{\mathbf{v}_2,\mathbf{v}_3,\mathbf{v}_4\}\nonumber\\
\mathbf{s}_3&=&\big(\mathbf{4}^{(m-1)},1^{\star}\big),\;\bm{\rho}_3=\{\mathbf{v}_4,\mathbf{v}_5\}\nonumber
\end{eqnarray}
The corresponding further splitting ratio table is
\begin{center}
\begin{tabular}{|c| c c c cc|}
\hline
 &  $\mathbf{v}_1$ & $\mathbf{v}_2$ & $\mathbf{v}_3$ &$\mathbf{v}_4$& $\mathbf{v}_5$ \\
\hline\hline
$\mathbf{s}_1$ & 9  & 8&&&   \\
\hline
$\mathbf{s}_2$ && 2& 2 &1& \\
 \hline
 $\mathbf{s}_3$ & &  & &1&0 \\
 \hline
 $\bm{\alpha}^{\rm LCM}$ &9& 8 &8&4& 0\\
 \hline
\end{tabular}
\end{center}
form which we obtain $\bm{\alpha}^{\rm LCM}=(9,8,8,4,0)$, implying both raw packet saving gain and further splitting ratio gain. Note that this LCM vector depends only on $\bar{t}$ sue to the selection of the set of transmitters.\hfill $\lozenge$
\end{example}

  \section{General Achievable Scheme of Theorem \ref{theorem_heter_pkt_size}}
\end{appendices}

\fi

\bibliographystyle{IEEEbib}
\bibliography{references_d2d}

\end{document}

%% file: macros.tex
\long\def\comment#1{}

\newfont{\bbb}{msbm10 scaled 700}

\newfont{\bb}{msbm10 scaled 1100}

\newcommand{\qv}{{\bf q}}

\newcommand{\sv}{{\bf s}}

\newcommand{\vv}{{\bf v}}

\newcommand{\Fm}{{\bf F}}

\newcommand{\Ac}{{\cal A}}

\newcommand{\Dc}{{\cal D}}

\newcommand{\Qc}{{\cal Q}}

\newcommand{\Sc}{{\cal S}}
\newcommand{\Tc}{{\cal T}}
\newcommand{\Uc}{{\cal U}}

\renewcommand{\arg}{{\hbox{arg}}}

\newcommand{\be}{\begin{equation}}
\newcommand{\ee}{\end{equation}}
\newcommand{\bea}{\begin{eqnarray}}
\newcommand{\eea}{\end{eqnarray}}

